\documentclass[onecolumn,usenatbib,useAMS,usegraphicx]{mn2e}

\usepackage{amsmath}
\usepackage{aas_macros}
\usepackage{longtable,lscape}
\usepackage{rotating}
\usepackage{multirow,multicol}
\usepackage{xspace}

\newcommand{\Msun}{{\rm M}_{\odot}}
\newcommand{\brems}{{bremsstrahlung}\xspace}
\newcommand{\Mvir}{{\rm M_{vir}}}
\newcommand{\Rvir}{{\rm R_{vir}}}
\newcommand{\Vvir}{{\rm V_{vir}}}
\newcommand{\Tvir}{{\rm T_{vir}}}
\newcommand{\LX}{{{\rm L_X}\,}}
\newcommand{\rhoc}{\rho_{\rm crit}}
\newcommand{\dd}{{\rm {d}}}
\newcommand{\DF}{{\small DF}\xspace}

\newcommand{\gtrsim}{\mathrel{\hbox{\rlap{\lower.55ex \hbox {$\sim$}}
                   \kern-.3em \raise.4ex \hbox{$>$}}}}
\newcommand{\lesssim}{\mathrel{\hbox{\rlap{\lower.55ex \hbox {$\sim$}}
                   \kern-.3em \raise.4ex \hbox{$<$}}}}

\title{Numerical Simulations of Hot Halo Gas in Galaxy Mergers}

\author[Manodeep Sinha, Kelly Holley-Bockelmann]
{Manodeep Sinha$^{1}$\thanks{E-mail: manodeep.sinha@vanderbilt.edu, k.holley@vanderbilt.edu}
, Kelly Holley-Bockelmann$^1$ \\
$^1$ Department of Physics \& Astronomy, Vanderbilt University  \\
}

\begin{document}
\maketitle

\begin{abstract}

Galaxy merger simulations have explored the behaviour of gas within the galactic disk, 
yet the dynamics of hot gas within the galaxy halo has been neglected. We report on 
the results of high-resolution hydrodynamic simulations of colliding galaxies with 
hot halo gas.  We explore a range of mass ratios, gas fractions and
orbital configurations to constrain the shocks and gas dynamics 
within the progenitor haloes. We find  that : {\it (a)} A strong shock  is produced in the 
galaxy haloes before the first passage, increasing the temperature of the gas 
by almost an order of magnitude to $T\sim 10^{6.3}$ K. {\it (b)} The X-ray luminosity 
of the shock is strongly dependent on the gas fraction; it is $\gtrsim 10^{39}$ erg/s for halo
gas fractions larger than 10\%. {\it (c)} The hot diffuse gas in the simulation produces
X-ray luminosities as large as $10^{42}$ erg/s. This contributes to the
total X-ray background in the Universe. {\it (d)} We find an analytic fit to the maximum X-ray 
luminosity of the shock as a function of merger parameters. This fit can be used in semi-analytic
recipes of galaxy formation to estimate the total X-ray emission from shocks
in merging galaxies. {\it (e)} $\sim$ 10-20\% of the initial gas 
mass is unbound from the galaxies for equal-mass mergers, while $3-5\%$ of the gas
mass is released for the 3:1 and 10:1 mergers. This unbound gas ends 
up far from the galaxy and can be a feasible mechanism to enrich the IGM with metals.

\end{abstract}


\section{INTRODUCTION}

Galaxy mergers drive galaxy evolution; they are one mechanism by which galaxies grow. 
The merger process is also accompanied by a strong gas inflow towards the central 
region \citep{BH91, MH94, C94, HM95, SM96} triggering a central starburst and perhaps an AGN.
Consequently, galaxy mergers have been studied quite extensively both numerically  \citep{H41,TT72,BH92, BH96, MH96,SW99, DMH99, B02}
and observationally  \citep{S86, R02, L03, CBG04}. However, none of these galaxy merger simulations included the
hot gas in the galactic halo that is expected from standard galaxy formation theory \citep{WR78,WF91}.
In this paper, we explore the effect of hot halo gas during galaxy mergers with high-resolution
numerical simulations.

Semi-analytic galaxy formation models have long predicted the existence of an
extended reservoir of hot halo gas around spiral galaxies \citep{WR78, WF91}.
In this scenario, gas falling into the dark matter potential shock-heats
to the virial temperature, $\Tvir = 10^6 {\rm K}\,(v_{\rm circ}/167 \,{\rm km/s})^2$
and subsequently cools over a characteristic time-scale, 
$t_c = k T/n_i \Lambda (T)$, that depends on the temperature $T$, the number density $n_i$
and the cooling function $\Lambda(T)$. This gas cools via thermal
\brems and atomic line emission processes, with the majority of
the radiation occurring in the soft X-ray band. The expected X-ray luminosity, $\LX$,
is a steep function of the circular velocity, $v_{\rm circ}$ --  $\LX \propto v_{\rm circ}^5$.  
Since the effect of mass (circular velocity) on this halo X-ray luminosity is so non-linear, 
any attempts to observe the X-ray signature of this gas would naturally be biased towards very massive haloes.
Consequently, hot gas in massive elliptical galaxies has been shown by X-ray observations 
of cluster galaxies as well as in isolated ellipticals \citep[see][and references
therein]{MB03}. Observations of massive ellipticals show that the X-ray luminosity
is non-stellar in nature \citep{OFP01} leading credence to its hot gaseous origin.
An extended gaseous halo in a quiescent
spiral has been observed {\em {only}} for the massive spiral NGC 5746 \citep{P06}.\footnotemark
However, the existence of a hot, rarefied and extended galactic halo has been inferred from 
OVI absorption lines in high velocity clouds in the vicinity of the Milky Way \citep{S03}.
The density of this halo gas ($n_{\rm H}  \sim 8\times 10^{-5} {\rm cm}^{-3}$) \citep{M00,S97}
is consistent with the predicted density for a Milky Way type galaxy.
Given the existence of hot halo gas, it is important to include it in galaxy merger
simulations.

\footnotetext{It has since been shown that the detection of hot gas in NGC 5746 may be
caused by contamination from unresolved point sources (J. Rasmussen, private communication). }

During a merger, the bulk motion of the galaxy gas is converted into thermal energy
via shocks. The orbital velocity of the gas during a merger is $\geq$ 100 km/s
whereas the sound speed is $\sim$ 10 km/s. This can produce strong shocks with Mach numbers $>10$. 
Assuming an adiabatic index of 5/3, the shocks may heat the gas to:
\begin{equation}
T_{\rm final} = \displaystyle\frac{3\,\mu {\rm m_H}}{16k}\, \mbox{v}_{\rm initial}^2\, ,
\label{eqn:rankinetemp}
\end{equation}
where T$_{\rm final}$ is the gas temperature after the shock, $\mu$ is the mean molecular
weight, ${\rm m_H}$ is the proton mass, $k$ is the Boltzmann constant, and v$_{\rm initial}$ is 
approximately equal to the radial velocity prior to the collision. Nearly 50\% of the kinetic 
energy can be converted into thermal energy of the 
gas particles. Thus, we expect that the hot gas will shock-heat and radiate in 
soft X-rays. The shocks may occur as soon as the gaseous haloes collide -- well before the 
optical extents of the galaxies begin to show signs of morphological disturbance.

This hot halo gas may play an important role in enriching the intergalactic medium (IGM). Galaxy
formation theory predicts that baryons fall into the dark matter potential and shock-heat. Absorption
lines reveal the presence of diffuse gas with column density $N_{\rm H} \gtrsim 10^{14}$/cm$^2$ \citep{FE82,CSKH95,W95}.
At $z \sim 3$, most of the baryonic mass is in the Lyman-$\alpha$ forest \citep*{FHP98}. 
The baryon fraction in the Lyman-$\alpha$ forest systematically decreases with $z$  and is almost zero at present. 
It would be reasonable to expect
that all the baryons are now part of galaxies in the form of stars and the interstellar medium. However,
recent estimates of the massfraction of baryons in different components still fail to 
detect $\sim 60\% $ of the total baryon content at low $z$, with $6\%$ in stars in galaxies \citep{FP04},
$2\%$ in neutral gas \citep{Z03}, $4\%$ in galaxy clusters and galaxy groups \citep{ASF02} and $29 \pm 4\%$ in 
low-$z$ Lyman-$\alpha$ absorbers \citep{PSS04,DS08}.
The consensus is that the majority of the `missing' gas is located in regions of low over-density, $\delta\rho/\rho \sim 10^0-10^2$ with
temperatures in the range $10^5-10^7$ K. This is the  Warm-Hot Intergalactic Medium (WHIM)
and is currently one of the most elusive astrophysical objects \citep[for a recent review see][]{B07}.  
Numerical simulations predict that the WHIM must reside in filaments and relatively low over-density
regions \citep{CO99,DHKW99,CO06,DO07}. Numerous efforts have been taken to detect and characterise the WHIM; 
however, most of the observations to date are 
not iron-clad and have required some fine-tuning in the fitting procedure 
for the data to come up with a significant detection \citep{NM05,MWC03}. The current instrumentation
limits on the {\it Chandra} and {\it XMM-Newton} are a major hold-up in the 
discovery of the WHIM; next generation X-ray telescopes like {\it Constellation-X} should be capable 
of detecting the absorption lines from the WHIM along the line of sight to bright background
sources.

The immediate question is: How does gas get into the WHIM state~? 
Some form of mass and  energy injection is essential to create this hot reservoir of gas,
and this form of feedback must also account for the metals in the IGM. Numerical
simulations of structure formation must incorporate various baryonic processes
such as supernova feedback \citep{MF99,BG02,TC00,ST06,GW07}, starburst-driven galactic winds \citep{SH03, OD08}
and AGN feedback \citep{SDH05,DSH05,SSB05,TSC06,HH06,CS06,SSMH07,BMK08,MC08}. 
All these energy delivery mechanisms are more effective
in low mass galaxies \citep{STWS08}, e.g., these prescriptions blow-out the gas in dwarf
galaxies \citep{MF99,KSW07}. {\em However, this is in direct disagreement with observational results that
show dwarf galaxies to be more gas-rich} \citep{K04}. In this paper, we will explore
if galaxy mergers, which are ubiquitous, can be used to fuel the IGM.

Observations of major galaxy mergers show that the merger rate evolves rapidly, increasing from 
$\sim$ 5-10\% at  $z\sim 1$ to $\sim$ 50\% at $z\sim 3$ \citep{CBDP03}. Integrating this merger rate
suggests that a typical massive galaxy will undergo 4-5 major mergers between
$z\sim 3$ and $z \sim 0$, with most of these mergers occurring at $z> 1.5$. Numerical simulations
can be used to explore the evolution of the merger rate as well as the merger ratio with $z$.
\citet{FM08} analyse the {\it Millenium Simulation} to constrain this evolution and find that
a major merger (merger ratio $\geq 1/3$) occurs at the rate of $0.5$ per halo per
unit redshift. Minor mergers with $\xi \geq 1/10$ are far more common at all redshifts. 

In this paper, we explore typical mergers with a Milky Way type primary galaxy.
We sample the major mergers by simulating 
an equal-mass merger between two such galaxies. We also attempt to get the boundary of the major 
mergers for a merger ratio, $\xi \sim 1/3$ for our second set of merger ratios. 
Minor mergers are the most frequently occurring mergers in the Universe; we model this by simulating a merger with
$\xi \sim 1/10$. This describes the three independent sets of simulations that we will perform
to constrain the effect of the merger ratio on the behaviour of the hot gas during galaxy
mergers. Mergers need an additional specification -- the merger orbit. We will describe
the parameters for specifying the content of the galaxies (see Sections~\ref{section:darkmatter} 
and \ref{section:gas}), and construction of the merger orbits in Section~\ref{section:mergerorbits}. 
In Section~\ref{section:results} we will present the results of our numerical simulations of 
galaxy mergers.  We will also derive an analytic fit for the peak
shock X-ray luminosity and the unbound gas mass as a function of the progenitor
galaxy masses and impact parameters. In Section~\ref{section:discussion} we discuss
the implications of our results and future work. 

\section{METHOD}\label{section:method}

We create an equilibrium model for an isolated galaxy by numerically sampling for the joint
distribution function (hereafter, \DF) of the dark matter and gas. The \DF for a 
spherically symmetric potential-density pair can be written as:
\begin{equation}
f(\mathcal{E}) = \displaystyle\frac{1}{\sqrt{8} \pi^2} \left [\displaystyle\int_{0}^{E}
     \displaystyle\frac{{\rm d}^2 \rho}{{\rm d} \Psi^2} \displaystyle\frac{{\rm d}
     \Psi}{\sqrt{\mathcal{E} - \Psi}} + \displaystyle\frac{1}{\sqrt{\mathcal{E}}} \Bigl(\displaystyle\frac{\dd \rho}{\dd \Psi}\Bigr)_{\Psi = 0}\right]
\label{eqn:nfwdf}
\end{equation}
where, $\mathcal{E}$ and $\Psi$ are the absolute value of the total energy and 
the gravitational potential respectively and $\rho$ is the density. We will describe
the initialisation of the dark matter and the gas component in the following subsections.

\subsection{Constructing a dark matter halo}\label{section:darkmatter}

Large scale cosmological simulations lead to a two-parameter universal dark matter halo density profile 
as shown by \citet*{NFW97} (hereafter, NFW). The key characteristics are the virial radius,  $\Rvir$ 
such that the average density of the halo within that radius is a certain multiple $(\Delta_{200} = 200)$ 
of the critical density of the universe, $\rhoc =3 H_0^2/8\pi G$, 
and the concentration parameter, $c=\Rvir/r_s$, where $r_s$ is the scale radius
of the NFW profile.  
Since the mass of the NFW halo is infinite, we implement an ad-hoc truncation radius
for the dark matter (and the gas) at $R_{\rm halo} = 1.2\times \Rvir$.

\begin{figure}
\centering
\includegraphics[scale=0.7,angle=-90,clip=true]{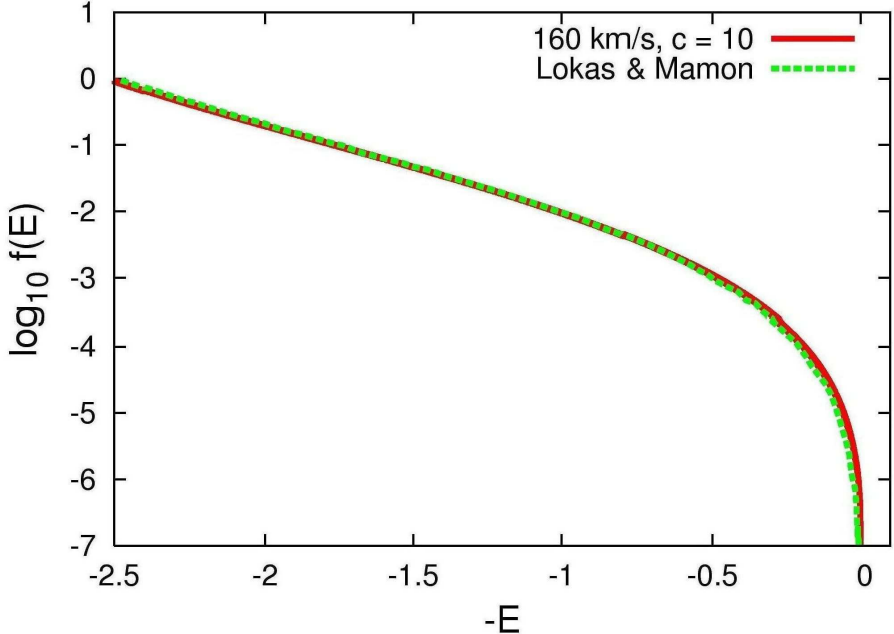}%
\\
\caption[Distribution functions for the Milky Way type halo]
{\small The computed \DF for our fiducial galaxy with $c=10$ and 10\% gas fraction.
Overplotted in magenta is the \DF from \protect\citet{LM01} for a dark matter only NFW halo with $\Vvir = 160$ km/sand $c=10$.
The presence of the gas makes very little difference in the overall \DF.}
\label{fig:nfwdf}
\end{figure}

To compute the \DF, we first create a logarithmically spaced array in $r$ with $10^5$ bins.
The minima and maxima of the bins are fixed at $10^{-5}$ and $100$ times the virial
radius. Thus, we capture the inner cusp of the NFW halo profiles as well as the
outlying particles with highly eccentric orbits; this can be seen from the
upturn in the \DF values for small $E$ in Fig.~\ref{fig:nfwdf}. Note that the inner limit
is much smaller than our softening length while the outer limit
is much larger than the truncation radius of $1.2$. This was done to ensure a higher
accuracy of the \DF itself, even though at present we can not reach such numerical 
resolution. On this finely-spaced grid, we define the values for $\rho(r)$, $M(r)$ and 
$\Phi(r)$ and obtain the derivatives by finite differencing. 
The \DF is then obtained by numerically integrating Eqn.~\ref{eqn:nfwdf} with a 
using a fourth-order Runge-Kutta integrator. Following \citet{LM01}, we
chose our system of units as $G=1.0, \,\Mvir=1.0, \, \Rvir=2.0$ with the resulting  $f(\mathcal{E})$ 
in units of $\sqrt{8}\Mvir/\left(\Rvir\,\Vvir\right)^3$.  

Our fiducial galaxy is a Milky-Way type galaxy with a circular velocity of $160$ km/s and a concentration
parameter $c=10$. $\Mvir$ for this primary galaxy is $1.36 \times 10^{12}\, \Msun$ with $\Rvir = 228.6$ kpc.
The secondary galaxies for our merger simulations are chosen to be one-third and one-tenth of the
mass of the primary; this translates approximately to circular velocities of $110$ and $74$ km/s. To determine
the concentrations of such haloes, we use the analytic fitting formula of \citet{BK01}.
With this prescription, our fiducial galaxy has a concentration of $c=10$, and the concentrations
for the two secondary galaxies are 16 and 25 respectively.

Fig.~\ref{fig:nfwdf} shows the comparison between the \DF  from \citet{LM01} for an
NFW halo with $c=10$ and the one obtained with our code for the fiducial galaxy with a 10\% gas fraction. 

\subsection{Constructing a gas-rich galaxy halo}\label{section:gas}

Now that the dark matter has been initialised, we need to initialise the gas. 
Earlier papers have used $\Omega_{\rm b}/\Omega_{\rm dm} \,\rho_{\rm dm}$ as the ad-hoc
density for the gas for cosmological simulations of structure formation. While that approximation
is valid for {\em {initial conditions}} in a cosmological framework where the force field is completely determined
by the dark matter (and gas merely falls into the dark matter potential), our simulations are in a fixed space-time
and we needed an analytic physically motivated gas density profile. We chose the observationally verified
$\beta$-profile \citep{CF76,JF84,ECFH98} used to model gas in galaxy clusters, a more massive analog to our gaseous haloes. The gas is
isothermal and in hydrostatic equilibrium. We use $\beta=2/3$ and a core of  $r_c = r_s/3$, giving
a density profile:
\begin{equation}
\begin{array}{lll}
\rho_{\rm gas} =  \displaystyle\frac{\rho_0}{1 + (r/r_c)^2}. \\
\end{array}
\label{eqn:betaprofile}
\end{equation}
Since the density profile is spherically symmetric,
we employ the Eddington inversion method to compute the \DF for the joint dark matter and gas distribution.
For the dark matter component, position-velocity pairs are drawn from this \DF. For the gas, we
chose the position from the gas density.
However, the gas velocity depends on the temperature and the temperature itself can not be freely
assigned since it is determined by the overall gas density-potential pair. 
We assume that the gas is polytropic and follows an equation of state, $P \propto \rho^{5/3}$ where $P$ and 
$\rho$ are the pressure and density of the gas. 
Finally, we assume hydrostatic equilibrium to compute the temperature of the
gas self-consistently. The gas is assumed to be at virial temperature at 
the virial radius with a central density constrained by the total amount of gas in the halo.
Once the temperature has been assigned to the gas particles, we calculate the velocity dispersion by assuming virialization.
The final gas velocities are drawn from a Gaussian distribution with zero mean velocity and a velocity dispersion given
by $\sigma^2 =  2\,k\,T/(\gamma-1)\mu\,m_H$.

The input parameters to our galaxy initialisation code 
are the circular velocity of the halo, the number of halo particles, the gas fraction and the
number of gas particles. Our fiducial galaxy is similar to the Milky Way with $\Vvir \,=\,160$ km/s. 
To investigate the role of mass-ratio during mergers, we constructed galaxies with $1/3$ and $1/10$
of the mass of our fiducial galaxy. For a galaxy with one-third the
mass, the virial velocity is $110$ km/s and for one-tenth the mass,
the virial velocity is $74$ km/s. Since we aim to simulate
the mergers of two different mass galaxies, it was imperative
to have the same mass  per particle between each of the galaxies to 
prevent spurious numerical relaxation. Thus,
the number of particles in the secondary galaxies are set by the corresponding number in the primary
galaxy. We use a fixed number of $5\times10^5$
particles for dark matter and gas in the primary galaxy for all the gas 
fractions. Since the total halo mass of the primary galaxy is fixed,
our halo mass resolution is $3\times10^6\,\Msun$, independent of the gas fraction.
The gas mass resolution for the 1\%, 10\% and 18\%  gas fractions in the primary
galaxy is $1.5\times10^6, \, 1.5\times10^7$ and $2.7\times10^7\, \Msun$  respectively (see Table~\ref{table:isolatedmodels}).\footnotemark

\footnotetext{Note that this gas mass resolution is the product
total number of neighbours of a gas particle, 50, and the mass of each gas
particle.}

\begin{figure}
\centering
\includegraphics[width=6.4in]{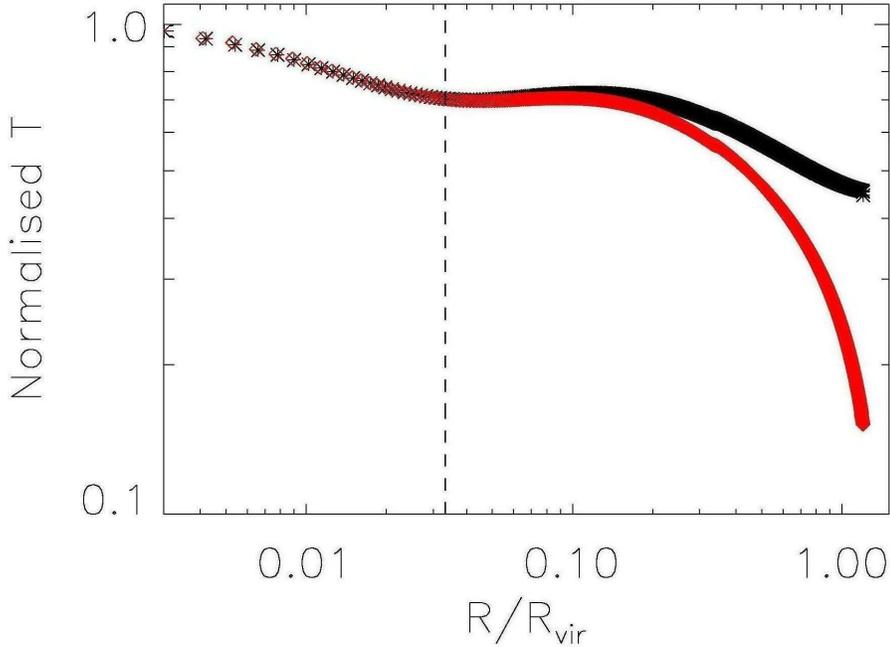}%
\caption[Iterations for a stable temperature profile]
{\small Creating a stable temperature profile. The black
shows the spherically averaged temperature profile obtained for the gas in hydrostatic equilibrium.
The red shows the temperature profile after an exponential drop-off has been added to the
outer region. The dashed line shows the core radius, $r_c = \Rvir/3\,c$
for the gas and corresponds to the location of the dip. The temperature
drops from $0.7\, \Tvir$ at $0.2 \,\Rvir$ to $0.4\, \Tvir$ at $0.6\, \Rvir$ and
is consistent with observations of galaxy clusters \protect\citep{VM05,LM08}.}
\label{figure:tempiter}
\end{figure}

\begin{figure}
\centering
\includegraphics[width=3.5in]{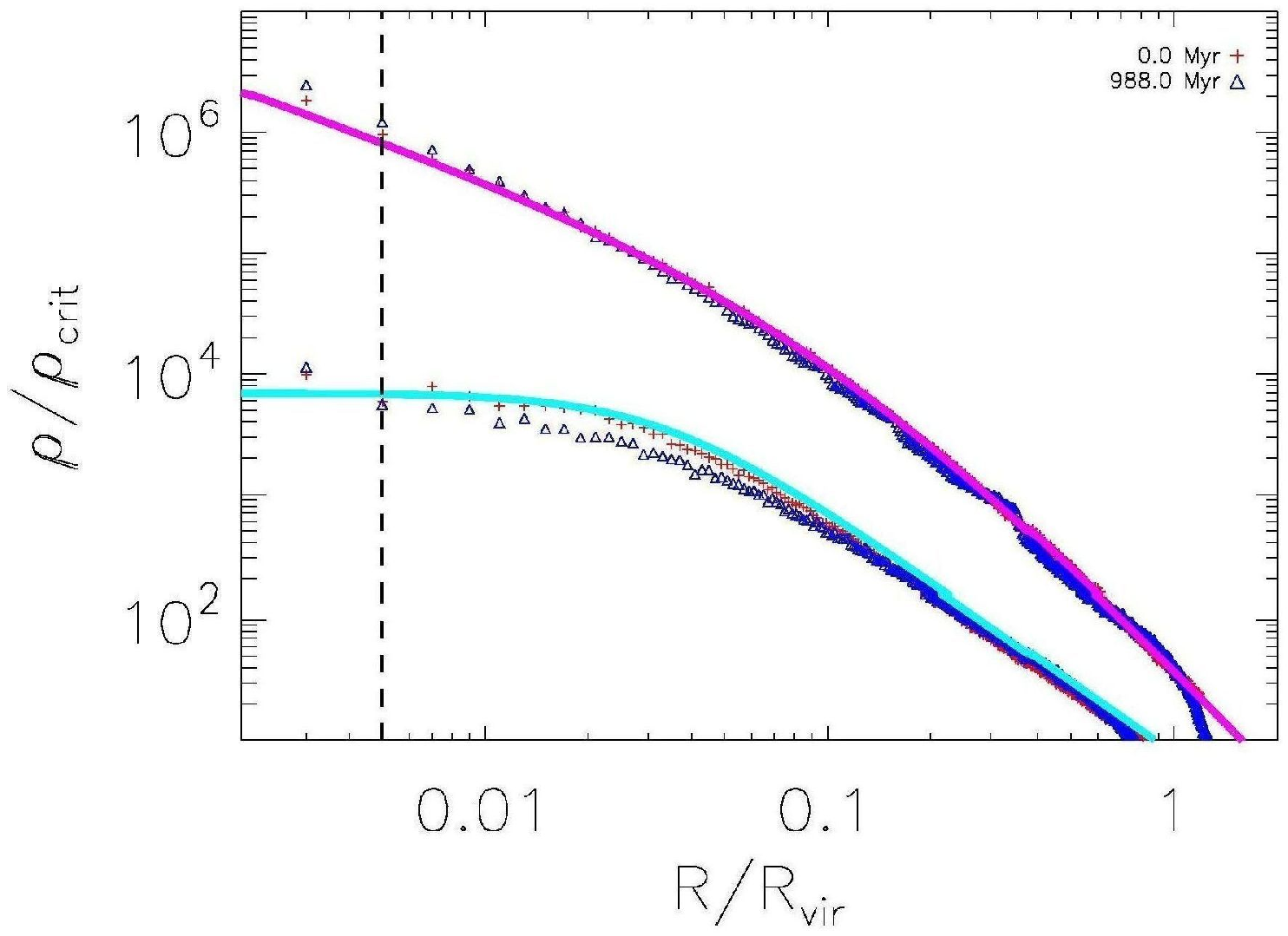}%
\includegraphics[width=3.5in]{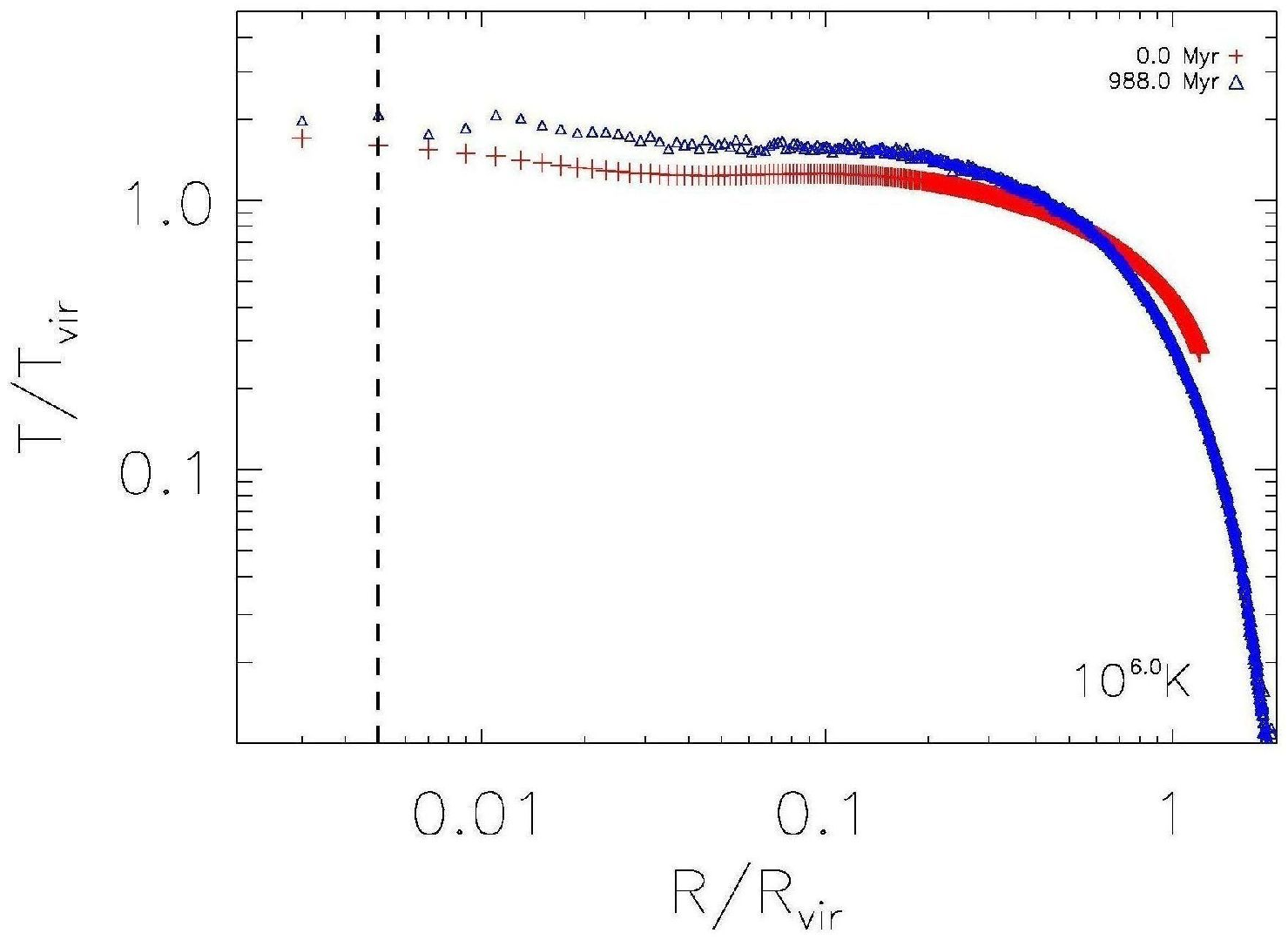}%
\\
\caption[Density and temperature profiles for 10\% primary galaxy]
{\small Evolution of gas density and temperature for a  10\% gas fraction in our fiducial
galaxy during initial condition construction. {\it Left}. The evolution for both the gas and dark matter is minimal; however,
we see some outflowing features in the dark matter density. This results in a
slightly larger galaxy at the end of 1 Gyr. The analytical input profiles shown in
magenta (for dark matter) and cyan (for gas). The vertical dashed line is the
softening used in the simulation.
{\it Right:} The virial temperature is noted in the bottom-right. As the system
evolves, the temperature profile drops off exponentially at roughly $\Rvir$ with 
a `knee' around $\sim 0.5\,\Rvir$.}
\label{figure:icprofiles}
\end{figure}

Since we are exploring how gas-rich galaxies behave during galaxy mergers, we did not want the input
temperature profile, $T(r)$, to vary and devised a multi-step process to reduce this temperature
evolution. Initially, the gas temperature is determined from the hydrostatic equilibrium condition in the
joint dark matter and gas potential. Then the gas velocity is assigned from the velocity dispersion
based on virial equilibrium. The problem is that this assumption is not valid in the inner
regions of the galaxy; even in isolation, the central gas particles shock-heat,
increase their temperatures and decrease their velocity dispersion rapidly. A higher artificial
viscosity parameter (AVP) aids this transformation and the core temperatures can increase by as much as a factor
of 2 from the originally solved temperature profile. To alleviate this problem, after initialising the galaxy with the 
method described above, we instead evolve the galaxy model in isolation with AVP of 0.5 to allow the 
gas particles to attain a new equilibrium temperature profile (see next section on the choice of AVP). 
This is the first iteration on an equilibrium
temperature profile for the halo gas. We find that the new temperature profile is indeed larger
at the centre and turns over at the outer edge with an exponential
drop-off near $r\, \gtrsim \,0.5\,\Rvir$. We fit this temperature
profile and apply it as a second iteration on the initial temperature profile for our model (see Fig.~\ref{figure:tempiter}
for a comparison of the two iterations). This evolved temperature profile is
nearly flat to $\sim 0.5 \Rvir$ and is consistent with observations of galaxy clusters showing that
the temperature stays to within 40\% of the central temperature out to $0.6\,\Rvir$ \citep{LM08,VM05}. \footnotemark
Finally, we evolve the galaxy with zero artificial viscosity over a period of $\sim$ 1 Gyr to get rid of
dynamical irregularities in our temperature profile fit (see Fig.~\ref{figure:icprofiles}). Since there is absolutely no entropy increase
with zero artificial viscosity, our temperature profile essentially remains constant in this step. 
\footnotetext{Since the surface brightness
decreases rapidly with radius, the signal in the outer regions is dominated
by low photon counts s and high background; consequently, the measurements are not as robust
in the outskirts of clusters. }
We re-calculate  $\Mvir$, $\Rvir$ and $\LX$ (see Table~\ref{table:isolated1}) 
after iterating the initial conditions. For the primary galaxy,
the maximum change in $\Rvir$ is $\sim 2.5\%$; for the one-third galaxy and
the one-tenth galaxy, the corresponding changes are $\sim 3\%$ and $\sim 2.6\%$
respectively. 
Once the galaxies are initialised, we check for stability 
by evolving them in isolation. One of the key
ways of ascertaining the equilibrium of multi-component systems
is by computing the {\it Virial of Clausius, VC}.
An equilibrium model for a galaxy should have $VC$ fairly close to 1. The
numerical models for our galaxies in the first iteration have a $VC$ $\sim 1.2$,
irrespective of the gas fraction or the halo size. Thus, while being close to equilibrium,
the galaxies are not exactly in steady state initially. The value of $VC$ for the isolated galaxies
with the iteratively-obtained temperature profile, ranges from 1.12 for the 
primary to $\sim 1.07$ for both the secondary galaxies. 

To test the stability of this new galaxy model after iterating the
temperature profile,  we simulate this evolved galaxy with AVP $=0.5$
for a Hubble time. We find that the change in the
overall temperature profile drops to $\sim 10-20\%$ over this
entire timespan. Our choice of softening length was guided by the suggested optimal softening from \citet{P03}
with $\epsilon_{\rm opt} \gtrsim 4\,\Rvir/\sqrt{N_{\rm{vir}}}$ where $N_{\rm{vir}}$
is the number of particles within the virial radius. Our $\epsilon = 1.1$ kpc is
comparable to the optimal value of $\epsilon_{\rm opt} \approx 0.9$ kpc.

We use the parallel, hydrodynamic code {\small GADGET-2} \citep{SYW01,S05}  for all the numerical 
simulations in this paper. {\small GADGET-2} computes the gravitational forces with a hierarchical 
tree algorithm \citep{BH86} while gas particles receive additional hydrodynamic acceleration 
as calculated using Smooth Particle Hydrodynamics \citep[hereafter SPH]{GM77}. The code explicitly 
conserves energy and entropy for the SPH particles and uses adaptive time-steps for the 
time-evolution. Since this is a first attempt to model the gas behaviour during the merger, cooling,
star formation and radiative feedback have all been neglected. 

\newpage
\renewcommand{\thefootnote}{\fnsymbol{footnote}}
\scriptsize
\begin{longtable}{ccccccccccc}
\caption[Input parameters for the isolated galaxy models]{This table lists
the parameters used to make the 9 isolated galaxy models. A constant 
physical softening of 1.14 kpc was used for both the gas and dark matter particles
in all simulations. The number of particles used to model the two smaller
galaxies was chosen such that the mass of each particle (dark matter and gas) was the same as
the mass of the corresponding particle for the primary galaxy. }
\label{table:isolatedmodels} \\

\hline \hline \\[-2ex]
   \multicolumn{1}{c}{\textbf{Gas Content}} &
   \multicolumn{1}{c}{\textbf{Galaxy Type}} &
   \multicolumn{1}{c}{\textbf{Conc. }} &
   \multicolumn{1}{c}{\textbf{v$\mathbf{_{\rm circ}}$}} &
   \multicolumn{1}{c}{\textbf{R$\mathbf{_{\rm vir}}$}} &
   \multicolumn{1}{c}{\textbf{M$\mathbf{_{\rm vir}}$}} &
   \multicolumn{1}{c}{\textbf{M$_{\rm DM}$}} &
   \multicolumn{1}{c}{\textbf{N$_{\rm DM}$}\footnotemark[2]} &
   \multicolumn{1}{c}{\textbf{Gas Mass Res.}\footnotemark[4]} \\[0.5ex]
   \multicolumn{1}{c}{\textbf{--}} &
   \multicolumn{1}{c}{\textbf{--}} &
   \multicolumn{1}{c}{\textbf{--}} &
   \multicolumn{1}{c}{\textbf{[km/s]}} &
   \multicolumn{1}{c}{\textbf{[kpc]}} &
   \multicolumn{1}{c}{$\mathbf{ [10^{10} \,\Msun ]}$} &
   \multicolumn{1}{c}{$\mathbf{ [10^{10} \,\Msun ]}$} &
   \multicolumn{1}{c}{\textbf[$\mathbf{10^5}$]} &
   \multicolumn{1}{c}{$\mathbf{ [10^{6} \,\Msun ]}$} \\\hline \hline \\[-2ex]
\endfirsthead

\multirow{3}{*}{1\% gas}
& Primary  & 10.0 & 160.0  & 228.6 & 136.1   &  150.0  & 5.0  & 1.5  \\
& Onethird & 16.0 & 110.0  & 158.5 &  45.3   &  48.0   & 1.6  & 1.5  \\
& Onetenth & 25.0 & 74.0   & 106.1 &  13.6   &  14.5   & 0.48  & 1.5 \\[0.7ex] \hline
\multirow{3}{*}{10\% gas}
& Primary  & 10.0 & 160.0  & 228.6 & 136.1   &  150.0  & 5.0  & 15.0 \\
& Onethird & 16.0 & 110.0  & 158.5 &  45.3   &  48.0   & 1.6  & 15.0 \\
& Onetenth & 25.0 & 74.0   & 106.1 &  13.6   &  14.5   & 0.48  & 15.0  \\[0.7ex] \hline
\multirow{3}{*}{18\% gas}
& Primary  & 10.0 & 160.0  & 228.6 & 136.1   &  150.0  & 5.0  & 27.0 \\
& Onethird & 16.0 & 110.0  & 158.5 &  45.3   &  48.0   & 1.6  & 27.0 \\
& Onetenth & 25.0 & 74.0   & 106.1 &  13.6   &  14.5   & 0.48  & 27.0 \\[0.7ex] \hline 
\hline \hline \\[-2ex]

\footnotetext[2]{The number of gas particles is the same as the number of halo particles in each galaxy.}
\footnotetext[4]{We used a fixed 50 SPH neighbours for all the simulations.}
\end{longtable}
\normalsize
\renewcommand{\thefootnote}{\arabic{footnote}}

\clearpage 
\begin{center}
\scriptsize
\begin{longtable}{cccccccc}
\caption[Evolution of isolated galaxies over 1 Gyr]{This table lists
some of the physical quantities for the galaxies after they were
evolved with zero artificial viscosity for 1 Gyr. The mergers 
are generated from two galaxies in this table.}
\label{table:isolated1} \\

\hline \hline \\[-2ex]
   \multicolumn{1}{c}{\textbf{Gas Content}} &
   \multicolumn{1}{c}{\textbf{Galaxy Type}} &
   \multicolumn{1}{c}{\textbf{Unbound gas }} &
   \multicolumn{1}{c}{\textbf{L$\mathbf{_{\rm X}}$}} &
   \multicolumn{1}{c}{\textbf{R$\mathbf{_{\rm vir}}$}} &
   \multicolumn{1}{c}{\textbf{M$\mathbf{_{\rm vir}}$}} &
   \multicolumn{1}{c}{\textbf{Hot gas}} &
   \multicolumn{1}{c}{\textbf{Virial of Clausius}} \\[0.5ex]
   \multicolumn{1}{c}{\textbf{--}} &
   \multicolumn{1}{c}{\textbf{--}} &
   \multicolumn{1}{c}{\textbf{[\%]}} &
   \multicolumn{1}{c}{\textbf{[$\mathbf{10^{40}}$ erg/s]}} &
   \multicolumn{1}{c}{\textbf{[kpc]}} &
   \multicolumn{1}{c}{$\mathbf{ [10^{10} \,\Msun ]}$} &
   \multicolumn{1}{c}{\textbf{[\%]}} &
   \multicolumn{1}{c}{\textbf{--}}  \\\hline \hline \\[-2ex]
\endfirsthead

\multirow{3}{*}{1\% gas}
& Primary  & 0.36 & 0.10  & 224.3 & 128.9  &  70.0  & 1.11  \\
& Onethird & 0.50 & 0.03  & 154.5 & 42.1   &  46.0  & 1.08  \\
& Onetenth & 0.58 & 0.01  & 103.9 & 12.8   &  21.0  & 1.07  \\[0.7ex] \hline 
\multirow{3}{*}{10\% gas}
& Primary   & 0.62 & 8.87  & 223.4 & 126.5  &  79.0  & 1.12 \\
& Onethird  & 0.43 & 3.17  & 153.9 & 41.6   &  48.0  & 1.08 \\
& Onetenth  & 0.56 & 0.88  & 103.7 & 12.8   &  24.0  & 1.06 \\[0.7ex] \hline 
\multirow{3}{*}{18\% gas}
& Primary   & 0.68 & 26.5  & 222.9 & 126.3  &  77.0  & 1.12 \\
& Onethird  & 0.42 & 10.9  & 153.6 & 41.3   &  50.0  & 1.08 \\
& Onetenth  & 0.50 & 2.75  & 103.3 & 12.6   &  25.0  & 1.06 \\[0.7ex]
\hline \hline \\[-2ex]
\end{longtable}
\normalsize
\end{center}

\subsection{Dependence on Artificial Viscosity}
\begin{figure}
\centering
\includegraphics[width=3.5in,clip=true]{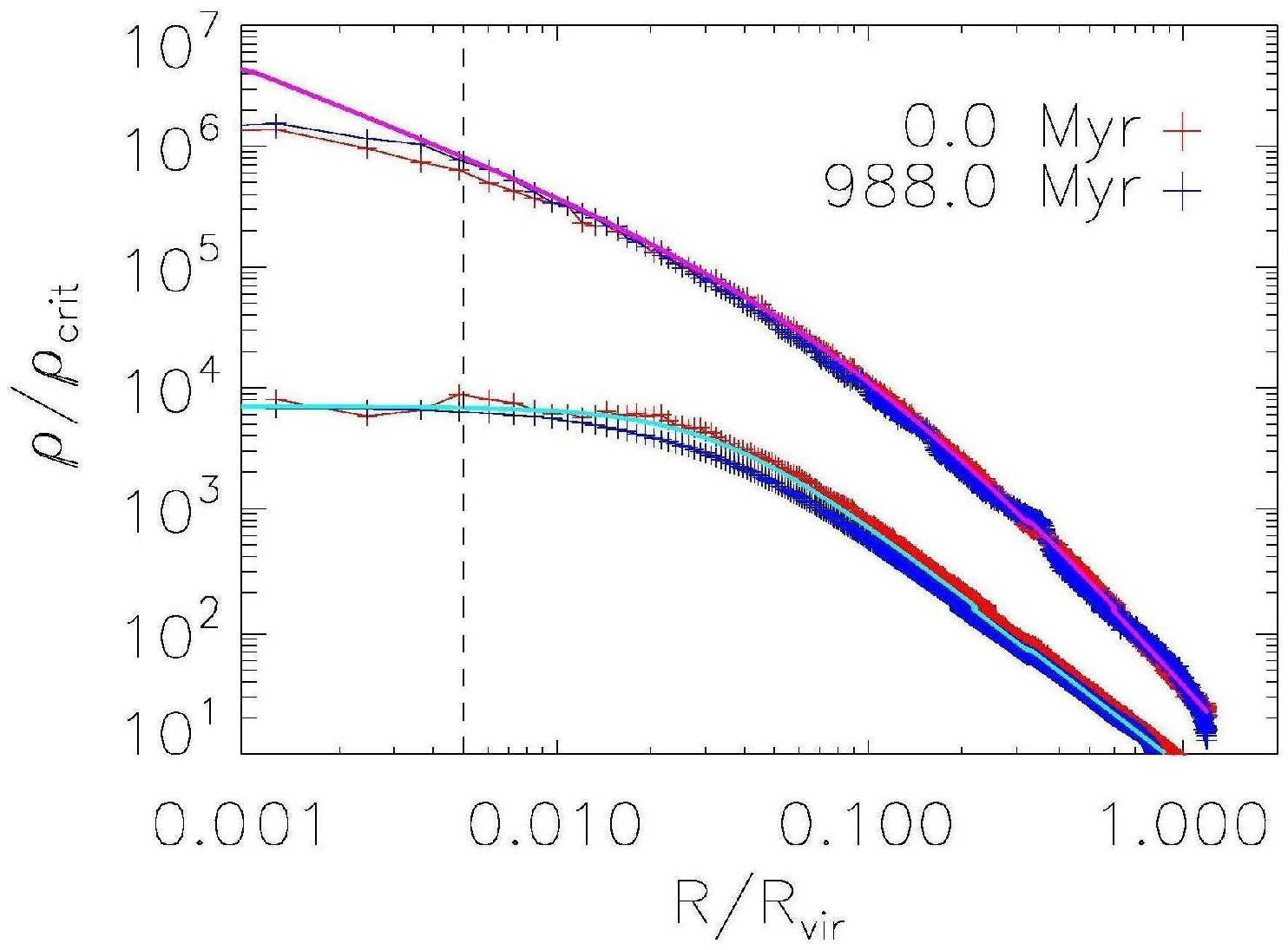}%
\includegraphics[width=3.5in,clip=true]{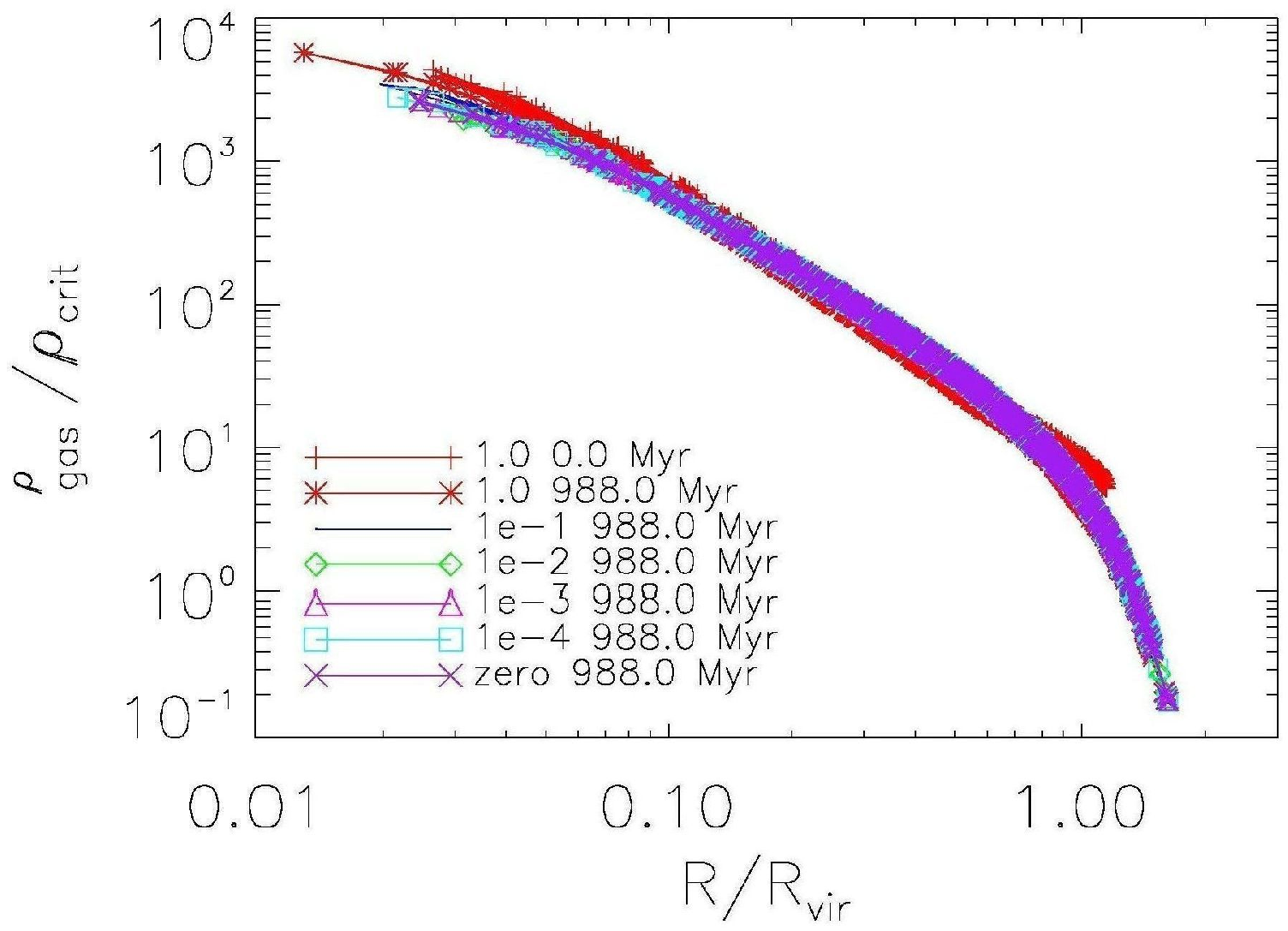}%
\\
\includegraphics[width=3.5in,clip=true]{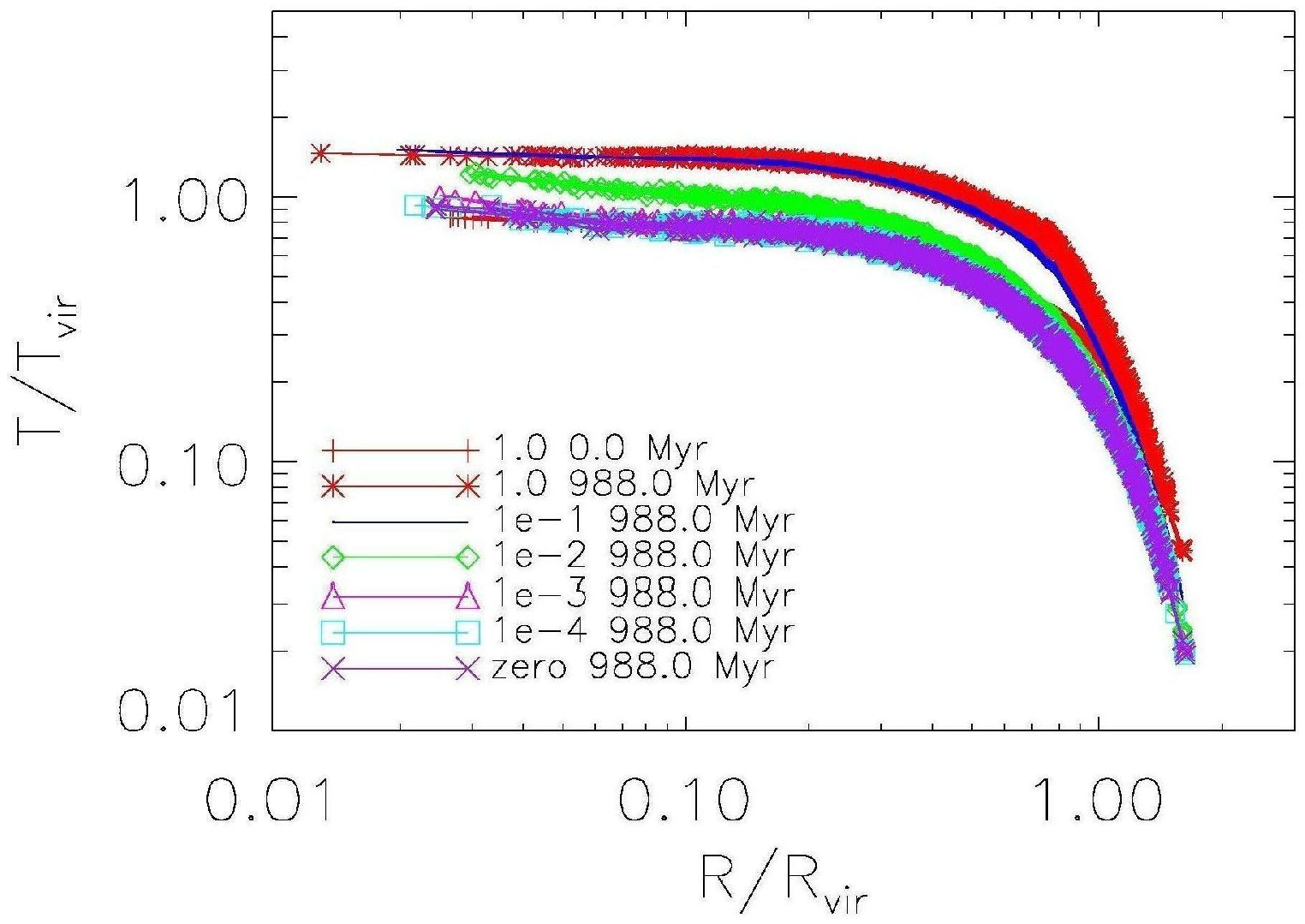}%
\includegraphics[width=3.5in,clip=true]{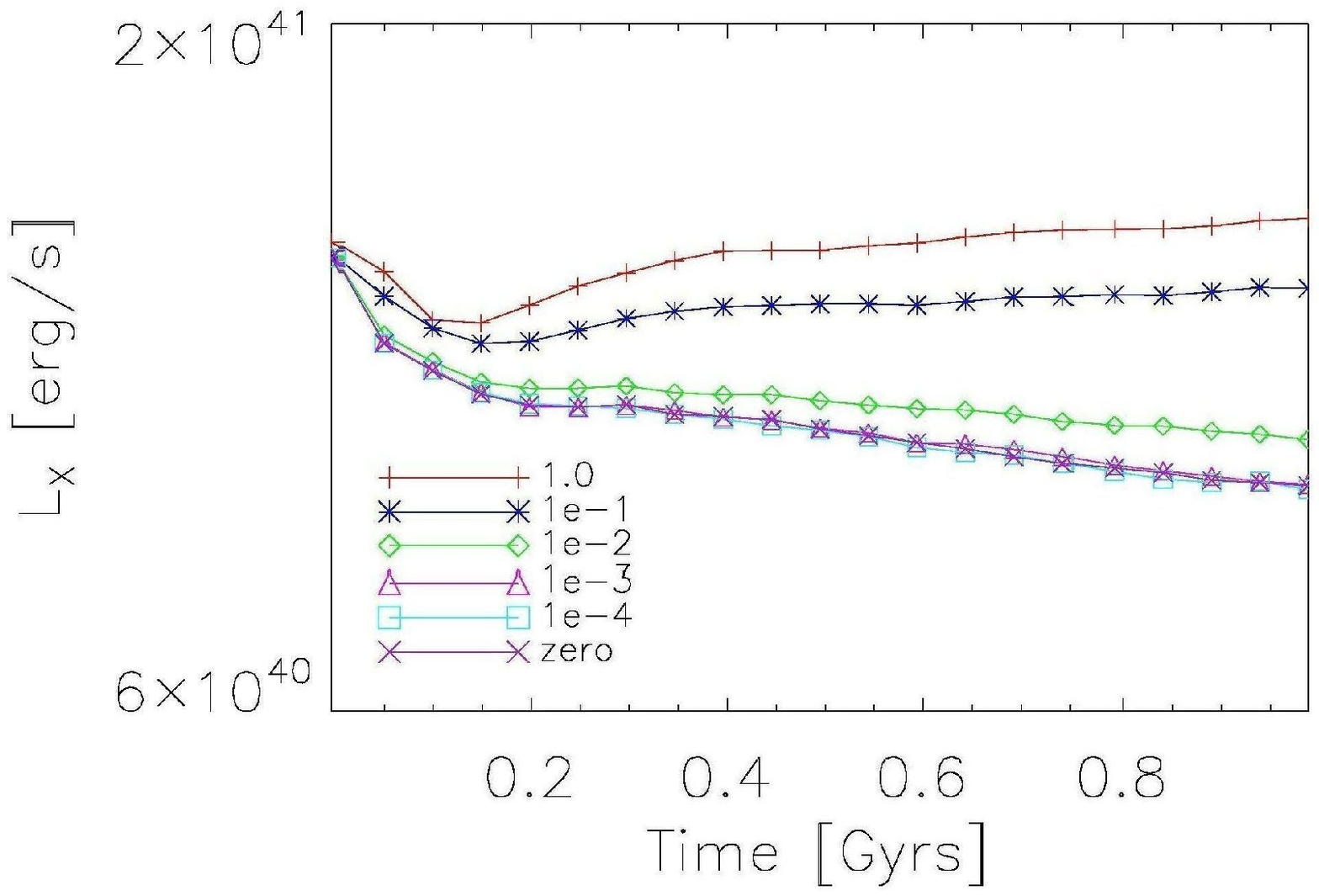}%
\\
\caption[Gas density and temperature evolution for different AVP]
{\small{ {\it Top Left}: The gas (lower curve) and dark matter (upper curve) density profile with the analytical $\beta$ and the
NFW profile overlaid as solid lines, for the primary galaxy with gas fraction of 10\%.
The galaxy is represented by $5\times10^5$ dark matter and gas particles each
and is evolved for 1 Gyr with an AVP$=0.1$. The flattening of the dark matter profile occurs in the inner
regions of the halo and corresponds to the softening  -- represented by the vertical dashed line.
{\it Top Right}: The same plot showing the gas density profile at 1 Gyr for different AVP's.
{\t Bottom Left}: The evolution of the temperature profile for  different AVP's. The
temperature increases by almost a factor of 2 for AVP $>0.01$.
{\it Bottom Right}: The evolution of the total X-ray emission via thermal
\brems from gas particles with $T > 10^{5.2} K$  and $\rho < 0.01 $ M$_\odot /pc^3$.}}
\label{fig:avptests}
\end{figure}

Since the AVP determines the strength of the shocks produced in the simulation,
we test the effect varying the AVP. We evolve the primary galaxy with a
10\% gas fraction in isolation, with varying AVP from $10^{-4}-1.0$.
Note that the recommended range of AVP is $0.1-1.0$ for numerical simulations. We find that the
difference in the final temperature profile is negligible between an AVP of 0.1
and 1.0; similarly for AVP's of $10^{-4}$ and 0. However, the average temperature
is  $20-30\%$ larger in the higher AVP case. Fig.~\ref{fig:avptests}
shows the evolution of density, temperature and $\LX$ from these set of simulations.
In keeping with the recommended value for AVP, we chose an AVP$=0.5$ to run all of our
merger simulations. We also ran an equal-mass merger with 10\% gas fraction and 0.1 $\Rvir$
impact parameter with three different AVP's of 0.1, 0.5 and 1.0. The results are
indistinguishable between the three, showing that we are in a convergent regime for this
physically reasonable value of AVP$=0.5$.

\subsection{Merger Orbits}\label{section:mergerorbits}

Cosmological simulations of large-scale structure formation shows that $\sim 40\%$ of the mergers occur 
with an eccentricity $e = 1.0 \pm 0.1$  and $\sim 85\%$  have an impact parameter $b > 0.1\,{\rm R_{vir}}$ 
\citep{KB06,B05}. More than $50\%$
of the mergers occur with $b > 0.5\,{\rm R_{vir}}$. Keeping these larger impact parameters in mind, we
create three orbits with $b= 0.01,\, 0.1 \,{\rm and} \, 0.5\, {\rm R_{vir}}$; -- two of our three orbits probe the larger
impact parameters. We chose elliptical orbits with $e\,\sim \,0.05,\;0.2\;{\rm \&}\;0.45$. We also have a suite of 
equal-mass	hyperbolic mergers.
Our simulations contain a set of 27 elliptical and 9 hyperbolic orbits, for a total of 36 simulations.
Table~\ref{table:mergersims} lists the eccentricity, total energy and the initial velocities for each
merger simulation.

\clearpage
\begin{center}
\renewcommand{\thefootnote}{\fnsymbol{footnote}}
\renewcommand{\arraystretch}{0.79}
\scriptsize
\begin{longtable}{rcrrrrr}
\caption[Table of orbital parameters for all the simulations]{The list of all the merger simulations
performed. The first three 1:1 mergers for each gas fraction are the hyperbolic
encounters, while the rest are all bound elliptical orbits with a fixed orbital
energy for a given merger ratio. }
\label{table:mergersims} \\

\hline \hline \\[-2ex]
   \multicolumn{1}{c}{\textbf{Gas}} &
   \multicolumn{1}{c}{\textbf{Merger Type}} &
   \multicolumn{1}{c}{\textbf{b}} &
   \multicolumn{1}{c}{\textbf{Eccentricity }} &
   \multicolumn{1}{c}{\textbf{Orbital Energy}} &
   \multicolumn{1}{c}{\textbf{Primary Vel.}} &
   \multicolumn{1}{c}{\textbf{Secondary Vel.}}  \\[0.8ex] 
   \multicolumn{1}{c}{\textbf{--}} &
   \multicolumn{1}{c}{\textbf{--}} &
   \multicolumn{1}{c}{\textbf{[kpc]}} &
   \multicolumn{1}{c}{\textbf{--}} &
   \multicolumn{1}{c}{\textbf{[$\mathbf{10^{56}}$ erg]}} &
   \multicolumn{1}{c}{\textbf{[km/s]}} &
   \multicolumn{1}{c}{\textbf{[km/s]}} \\[0.5ex]\hline \hline \\[-2ex]
\endfirsthead

\multirow{12}{*}{\begin{sideways}{\Large{1\% gas}} \end{sideways}}
& 1:1  & 2.3   & 2.99 & 2864.6  & 137.8 &  137.8 \\
& 1:1  & 22.8  & 2.95 & 2864.6  & 137.8 &  137.8 \\
& 1:1  & 114.3 & 2.78 & 2864.6  & 137.8 &  137.8 \\
& 1:1  & 2.3   & 0.06 & -1432.3 & 68.9  &  68.9 \\
& 1:1  & 22.8  & 0.18 & -1432.3 & 68.9  &  68.9\\
& 1:1  & 114.3 & 0.41 & -1432.3 & 68.9  & 68.9 \\
& 3:1  & 1.9   & 0.06 & -542.0  & 29.5  & 92.2\\
& 3:1  & 19.4  & 0.20 & -542.0  & 29.5  & 92.2\\
& 3:1  & 96.8  & 0.44 & -542.0  & 29.5  & 92.2\\
& 10:1 & 1.7   & 0.07 & -188.5  & 10.5  & 108.9\\
& 10:1 & 16.7  & 0.21 & -188.5  & 10.5  & 108.9\\
& 10:1 & 83.7  & 0.48 & -188.5  & 10.5  & 108.9\\[0.1ex]\hline\\[0.1ex]
\multirow{12}{*}{\begin{sideways}{\Large{10\% gas}} \end{sideways}}
& 1:1   & 2.3   & 2.99 & 3397.9  & 143.9 & 143.9\\
& 1:1   & 22.8  & 2.95 & 3397.9  & 143.9 & 143.9\\
& 1:1   & 114.3 & 2.78 & 3397.9  & 143.9 & 143.9\\
& 1:1   & 2.3   & 0.05 & -1698.9 & 71.9  & 71.9 \\
& 1:1   & 22.8  & 0.18 & -1698.9 & 71.9  & 71.9 \\
& 1:1   & 114.3 & 0.41 & -1698.9 & 71.9  & 71.9 \\
& 3:1   & 1.9   & 0.06 & -642.9  & 30.8  & 96.3\\
& 3:1   & 19.4  & 0.20 & -642.9  & 30.8  & 96.3\\
& 3:1   & 96.8  & 0.44 & -642.9  & 30.8  & 96.3\\
& 10:1  & 1.7   & 0.07 & -223.6  & 10.9  & 113.6\\
& 10:1  & 16.7  & 0.21 & -223.6  & 10.9  & 113.6\\
& 10:1  & 83.7  & 0.48 & -223.6  & 10.9  & 113.6\\[0.1ex]\hline\\[0.1ex]
\multirow{12}{*}{\begin{sideways}{\Large{18\% gas}} \end{sideways}}
& 1:1   & 2.3   & 2.99 & 3910.1  & 149.0 & 149.0\\*
& 1:1   & 22.8  & 2.95 & 3910.1  & 149.0 & 149.0\\*
& 1:1   & 114.3 & 2.78 & 3910.1  & 149.0 & 149.0\\*
& 1:1   & 2.3   & 0.05 & -1955.1 & 74.5  & 74.5\\*
& 1:1   & 22.8  & 0.18 & -1955.1 & 74.5  & 74.5\\*
& 1:1   & 114.3 & 0.41 & -1955.1 & 74.5  & 74.5\\*
& 3:1   & 1.9   & 0.06 & -739.8  & 31.9  & 99.7\\*
& 3:1   & 19.4  & 0.20 & -739.8  & 31.9  & 99.7\\*
& 3:1   & 96.8  & 0.44 & -739.8  & 31.9  & 99.7\\*
& 10:1  & 1.7   & 0.07 & -257.3  & 11.3  & 117.6\\*
& 10:1  & 16.7  & 0.21 & -257.3  & 11.3  & 117.6\\*
& 10:1  & 83.7  & 0.48 & -257.3  & 11.3  & 117.6\\*[0.3ex]\hline \hline \\[-2ex]
\end{longtable}
\normalsize
\renewcommand{\thefootnote}{\arabic{footnote}}
\renewcommand{\arraystretch}{1.0}
\end{center}

\section{Analysis}\label{section:analysis}
\subsection{Calculating X-ray emission}
We assume that the primary source of X-rays in a galaxy merger is thermal \brems
from hot, diffuse gas of primordial composition. In real galaxies, X-rays are also produced by
discrete sources like supernova remnants, compact binary objects as well as accretion
events occurring around massive black holes. However, such emission is more likely to
occur in the denser central regions of the galaxy and hence, also more likely to be
obscured by the larger column density of the intervening gas and dust. The hot, diffuse
gas should be relatively less affected by in situ absorption \citep{C06}.
Following \citet{C06}, we identify only those particles that have $T > 10^{5.2} K$ and density $\lesssim 0.01 \,\Msun\, {\rm pc}^{-3}$
as X-ray emitting hot, diffuse gas. Throughout this paper, we will use the term hot gas to denote those
gas particles that satisfy this criteria. Therefore, we estimate the X-ray luminosity  as:
\begin{equation}
L_{\rm X} = 1.2 \times 10^{-24}\, (\mu m_{\rm p})^{-2}\, \sum_i m_i\, \rho_i\, T_i^{1/2} {\rm erg \, s}^{-1},
\label{eqn:emission}
\end{equation}
where $m_{\rm p}$ is the mass of a proton, $m_i, \rho_i, T_i$ is the mass, density and temperature  of a gas particle
respectively and the summation is over only the hot gas particles.\footnotemark
\footnotetext{Note that enriched gas at $\sim 10^6$ K will cool primarily through metal line emission,
implying that the X-ray emission could well be enhanced in actual colliding galaxies.}

The orbital energy in the gas is $\gtrsim  10^{58}$ erg; much of this energy gets redistributed as the
internal kinetic energy of the remnant. The average X-ray luminosity in all the simulations
is $\sim 10^{41}$ erg/s and it would take  more than 10 Gyrs to radiate away only the
orbital energy via \brems. To ensure that we select only those gas particles that will remain
hot, we compute the cooling times for the hot gas particles and reject the ones that
have cooling time less than the dynamical time. As a further safeguard, we also reject the
hot gas particles that have a cooling time less than the time between consecutive
snapshots ($\sim 100 $ Myrs for all simulations). After these two step rejection
process we recompute the X-ray emission from the hot gas and the shocked gas
particles and find that the X-ray emission changes by less than 5\%  for the hot
gas and less than 1\% for the shocked gas particles. Thus, our predictions regarding 
X-ray emission from the hot and the shocked gas are not susceptible to significant changes
from radiative losses. 

\subsection{Shock Detection}

In our simulations, large-scale shocks develop in the gas. Since our simulations
are adiabatic, shocks are the only mechanism that can increase the entropy
of a gas particle. We take advantage of the conservative-entropy formulation \citep{SH02} of
{\small GADGET-2} \citep{S05} to identify the shocked particles based on their
rate of entropy change. Each snapshot file in {\small GADGET-2} 
gives the rate of change of entropy for every gas particle. We set a threshold for 
identifying shocked gas particles. We chose the threshold conservatively, i.e., we
are likely to miss some shocked particles. The threshold, in code units, is 
is $10^{10}$, $3\times10^9$ and $2\times10^9$ for 1\%, 10\% and 18\% gas fractions respectively.
We also locate every particle that was ever shocked by analysing all
the snapshots and tracking the particle ids. This helps to determine the mechanism behind
a particle's unbinding.

\subsection{Unbound particles}
At the beginning of the simulation, we identify the most bound 10\% of the particles 
of each galaxy and track them  to follow the centre of mass. The centre of mass velocity
is subtracted to compute the total energy of individual particles. 
Particles that have a total energy greater than zero are considered unbound from the system. By defining the unbound 
material in this fashion, we avoid tagging particles that could be unbound from each galaxy but not
unbound from the entire system. In the simulations where
we have a merger remnant, we identify the unbound particles at the end of the simulation (when the system has
settled down) and track those particles through the duration of the merger.

\section{RESULTS}\label{section:results}

In this section we explore how different merger ratios, gas fractions and impact parameters
affect the X-ray emission and the unbound gas. A priori, we expect to find an increasing trend
for X-ray emission with increasing progenitor masses --  this is due to the higher
virial temperature and larger hot gas fraction in the more massive haloes (see 
Table~\ref{table:isolated2}). In addition, a smaller impact parameter implies denser parts of 
the galaxy inter-penetrate; this should lead to stronger shocks and larger hot gas fractions. Therefore we expect
the X-ray emission from both the regular hot gas component and the shocked 
component to increase with a decreasing impact parameter. A larger impulse
is delivered during a close encounter with more massive galaxies; therefore,
we expect an increasing trend in the unbound gas fractions with increasing
galaxy mass and decreasing impact parameter. Thus, both $\LX$ and the unbound
gas fraction should behave in an analogous manner; we  test these hypotheses in this section.

\subsection{X-ray emission}

\begin{figure}
\centering
\includegraphics[scale=0.55,clip=true,bb=0 0 743 572]{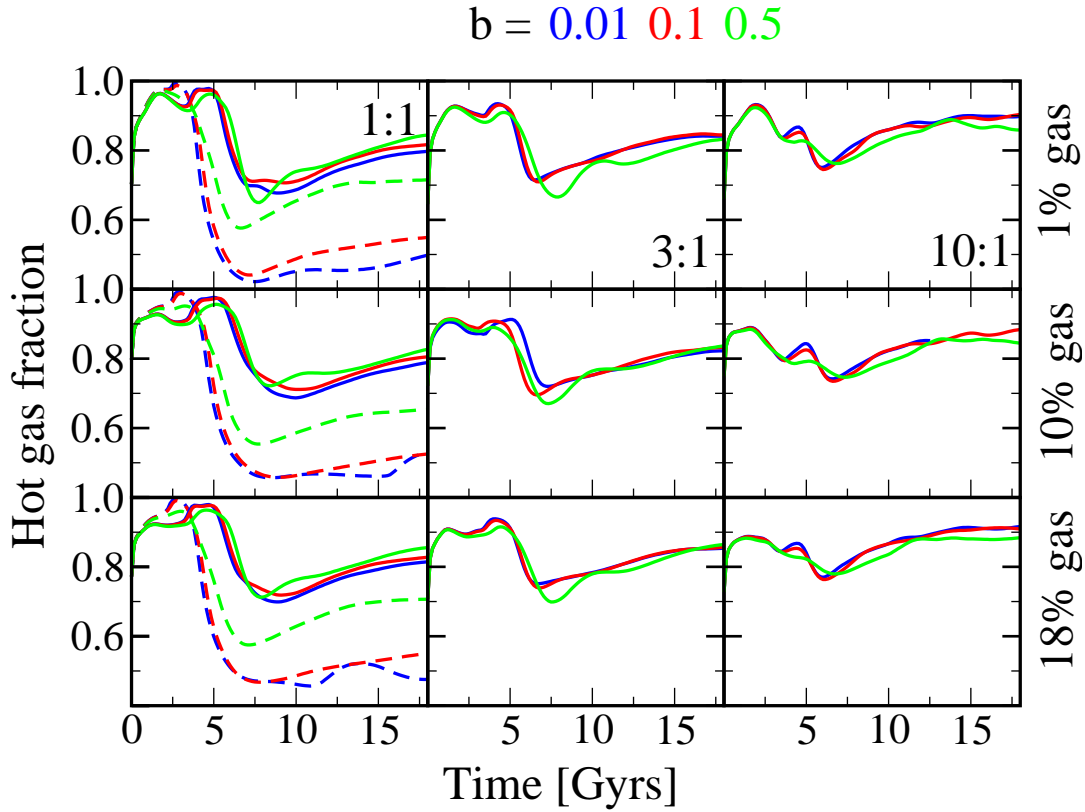}
\caption[Hot gas fraction for all simulations]
{\small The evolution of the hot gas fraction for all the 36 simulations. The left column contains 1:1 mergers,
the middle column contains  3:1 mergers and the right column contains the
10:1 mergers. The top row, middle and the bottom row show simulations
with 1\%, 10\% and 18\% gas fractions respectively. 
The blue, red and the green shows the 0.01, 0.1 \& 0.5 $\Rvir$ impact 
parameters respectively. The dashed lines in the left column show the
hyperbolic 1:1 mergers. 
We see that the peaks of the hot gas fractions corresponds
to the pericenter passages. The most drastic evolution is seen in the equal-mass mergers
where hot gas fraction evolves from $\sim 80\%$ to $\sim 50\%$ for 0.01 and 0.1 $\Rvir$.
For all the other simulations the final hot gas fraction reaches equilibrium at around
$80\%$.}
\label{fig:allsimshotgasfrac}
\end{figure}

\begin{figure}
\centering
\includegraphics[scale=0.55,clip=true,bb=0 0 743 572]{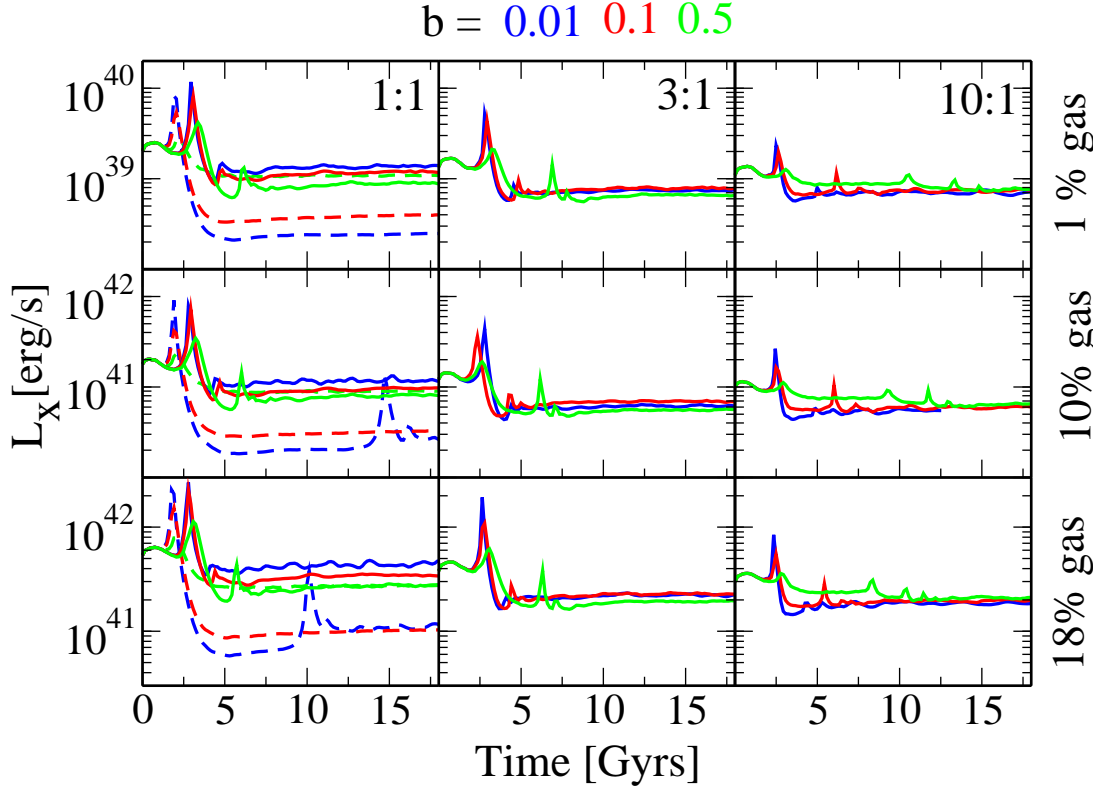}
\caption[L$_X$ for all simulations]
{\small The X-ray luminosity versus time for all simulations. 
The peak X-ray luminosity increases with increasing gas fraction $\propto \rho_{\rm gas}^2$ 
and occurs at the first pericentre pass. The peak X-ray
luminosity also shows an increasing trend with decreasing impact parameter and decreasing secondary
galaxy mass, correlated with the amount of shock heating that occurs. The line types and the colours
are the same as in Fig~\ref{fig:allsimshotgasfrac}.}
\label{fig:allsimslx}
\end{figure}

Though the isolated galaxies have varying initial hot gas fractions, during the course of the merger
more than 90\% of the gas is in the hot phase. This is caused primarily by  shock heating
during the encounter. For example, the smallest secondary galaxies have $\sim 40\%$ of their gas in
the hot phase in isolation (see Table~\ref{table:isolated2}), but that fraction increases to 
more than 90\% during the merger process. The total X-ray luminosity of the merger is determined by this
hot gas fraction and can be seen from the correlated decrease in both the hot gas fraction and the X-ray 
luminosity (see Fig.~\ref{fig:allsimshotgasfrac} and Fig.~\ref{fig:allsimslx}). 

For equal-mass mergers, the elliptical and hyperbolic orbits produce a peak $\LX$ of $\sim 10^{40}$ erg/s for the 1\% gas
fraction. This is from the first pericentre pass, which creates the strongest compressive
forces. Since $\LX$ is proportional to the square of the density, the shocks at this stage create the largest
X-ray emission. However, note that even though we are talking about close passages, the average
distance between the centres is greater than 20 kpc, greater than the typical disk size in spiral galaxies. 
Hence, the shocks will be observed at large distances from the centres. 
For equal-mass mergers with 10\% gas, $\LX$ peaks at $\sim 8\times 10^{41}$ erg/s during
pericenter pass. The 18\% gas fraction simulations produce $\LX$ as high as $\sim 2\times10^{42}$ erg/s. 

The simulations with the smallest impact parameters cause the denser parts to collide. 
As the impact parameter increases to 0.5 $\Rvir$, the lower density regions of the galaxy inter-penetrate. Since our simulations
are adiabatic, the maximum increase in the density can be a factor of 4; the strong dependence of $\LX$
on the gas density results in a lower $\LX$. 

\begin{figure}
\centering
\includegraphics[width=3.2in,clip=true,bb= 0 0 693 520]{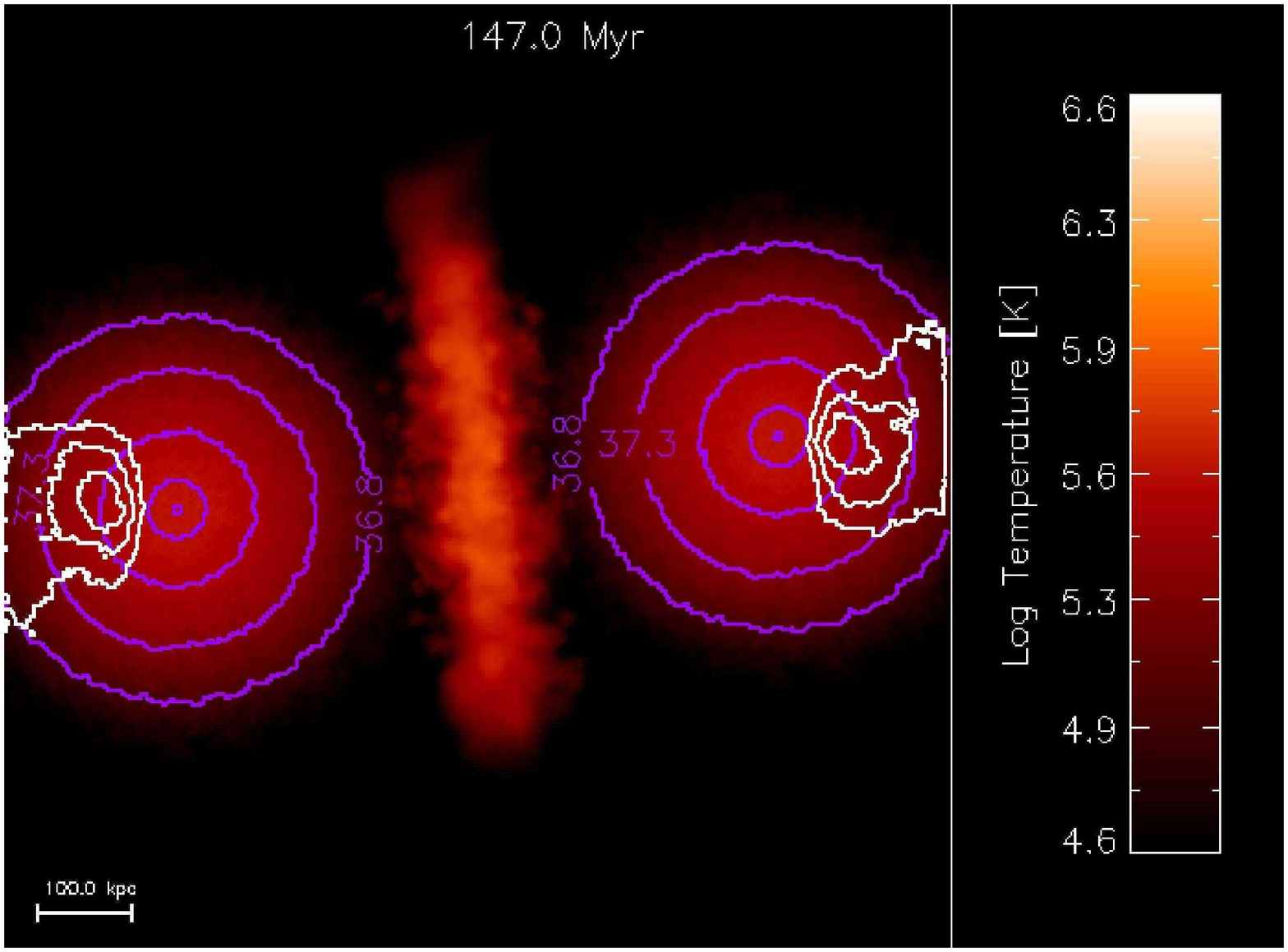}%
\includegraphics[width=3.2in,clip=true,bb=0 0 693 520]{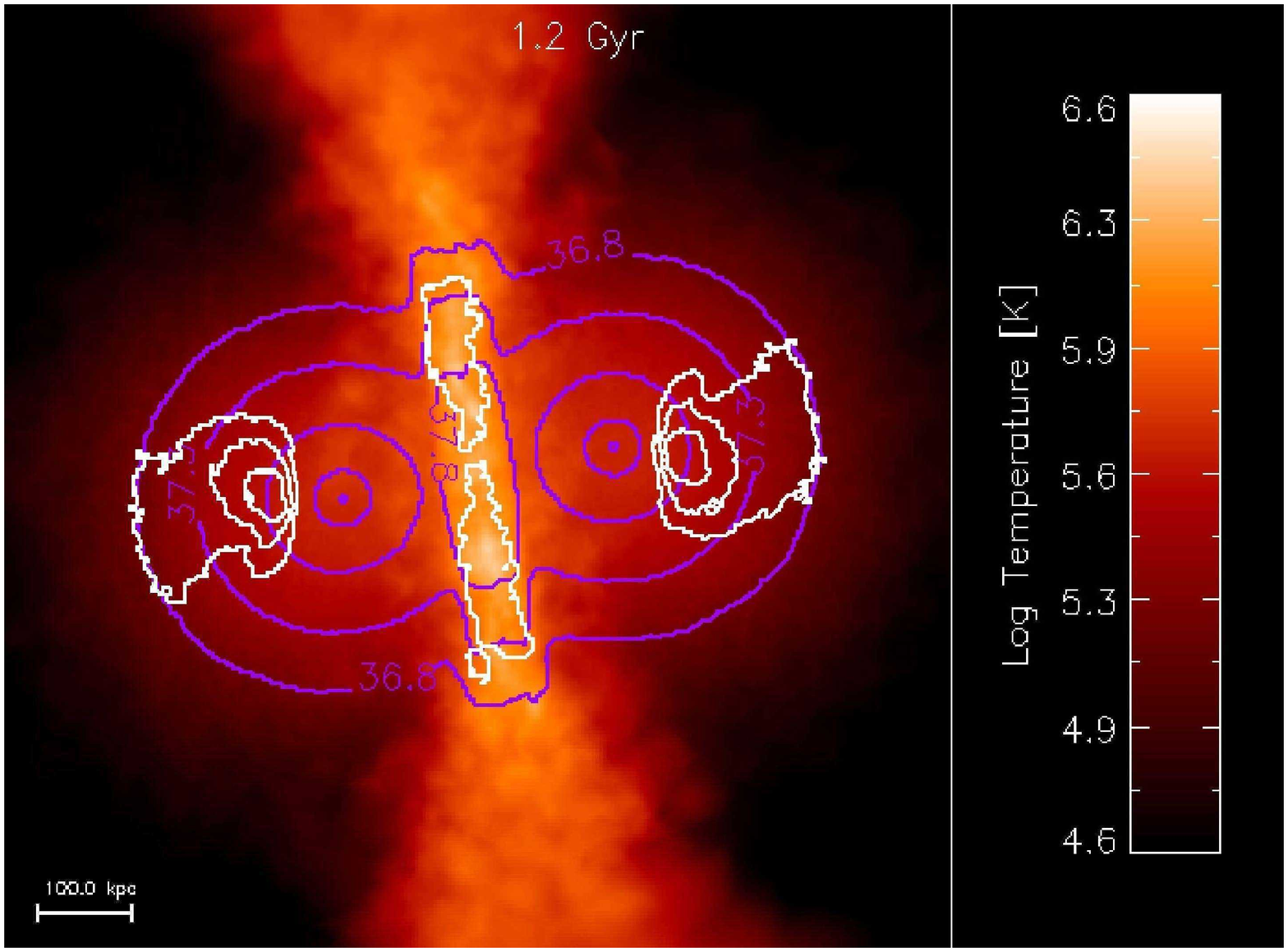}%
\\
\includegraphics[width=3.2in,clip=true,bb=0 0 693 520]{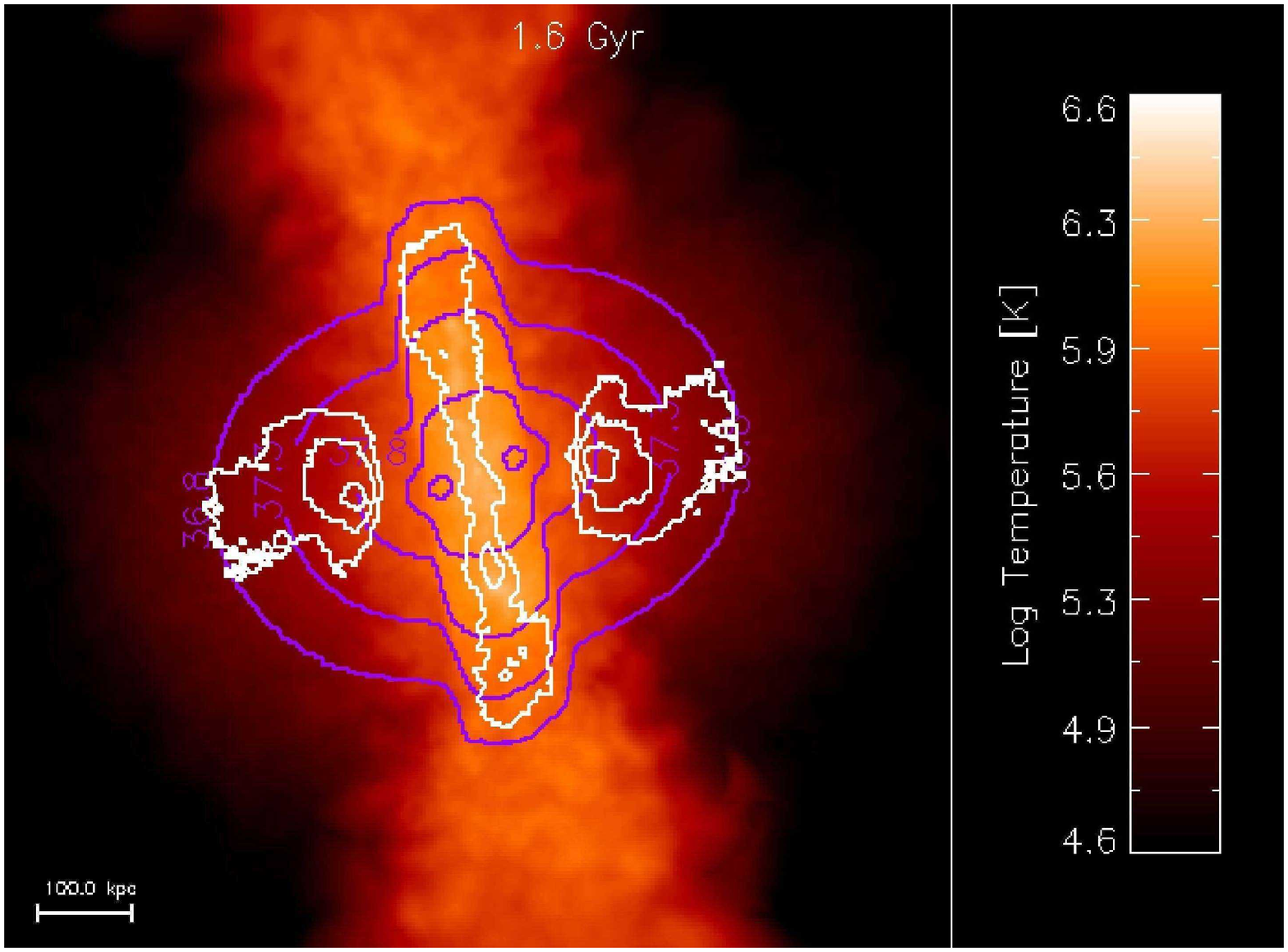}%
\includegraphics[width=3.2in,clip=true,bb=0 0 693 520]{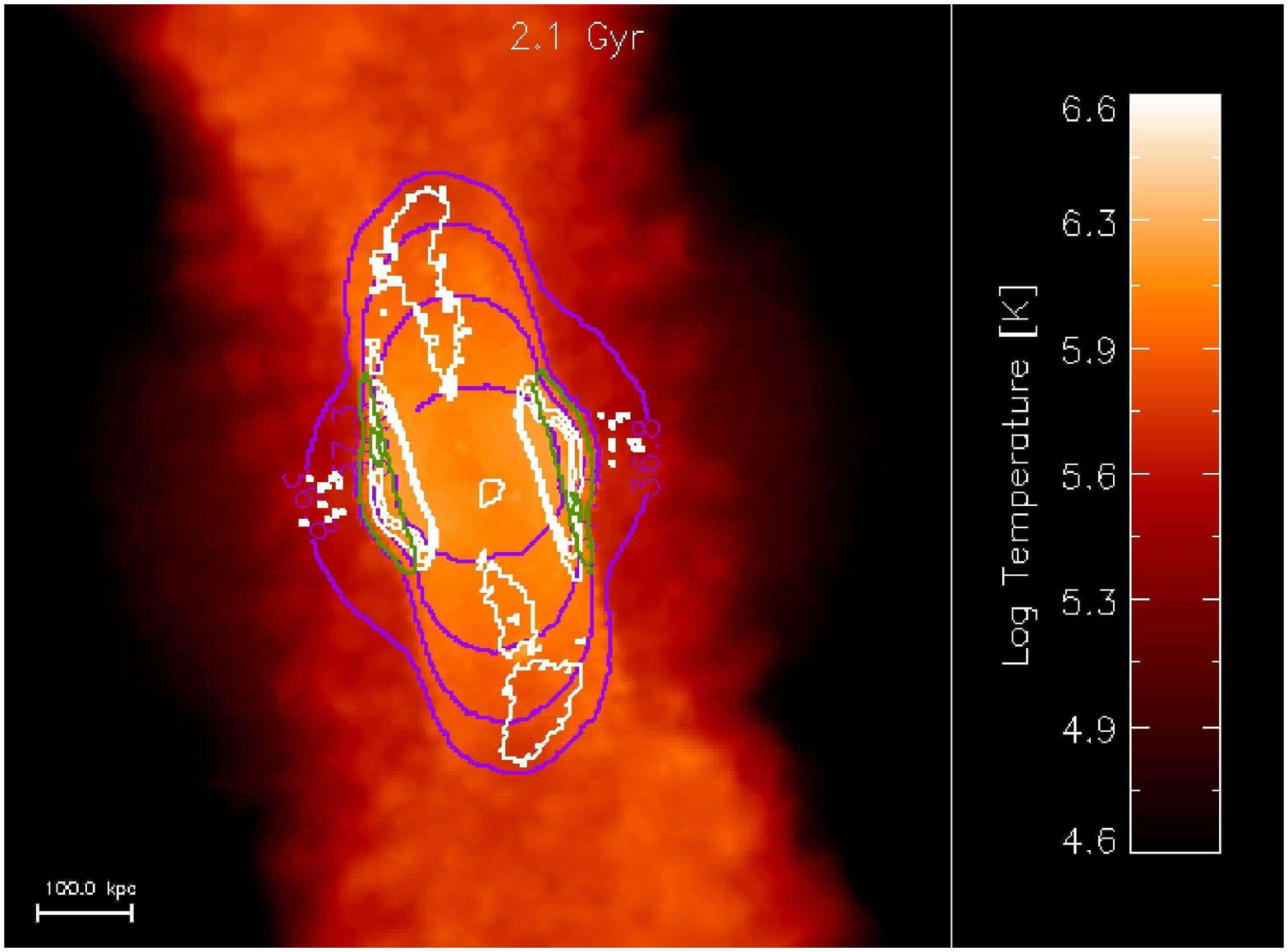}%
\\
\includegraphics[width=3.2in,clip=true,bb=0 0 693 520]{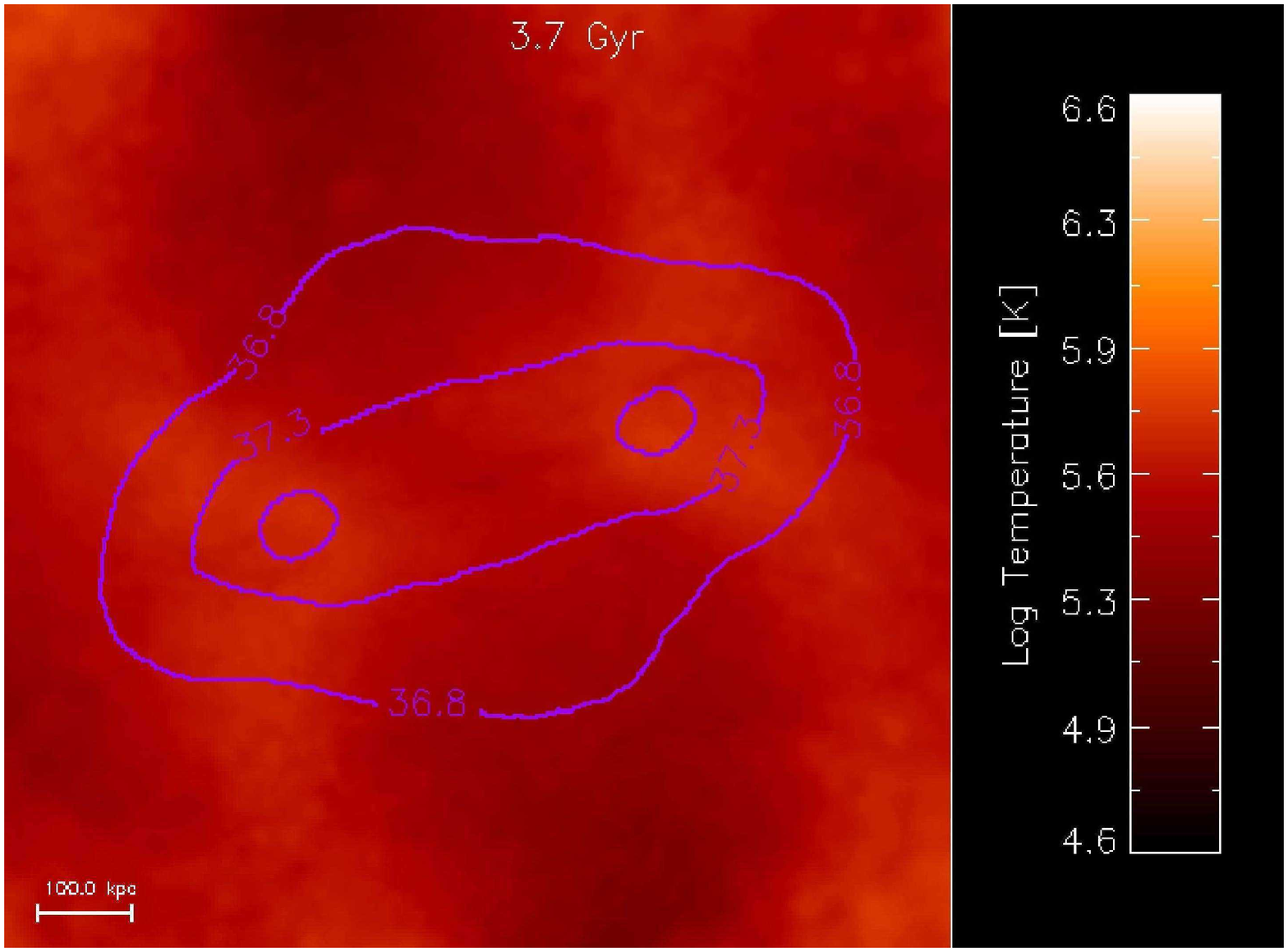}%
\includegraphics[width=3.2in,clip=true,bb=0 0 693 520]{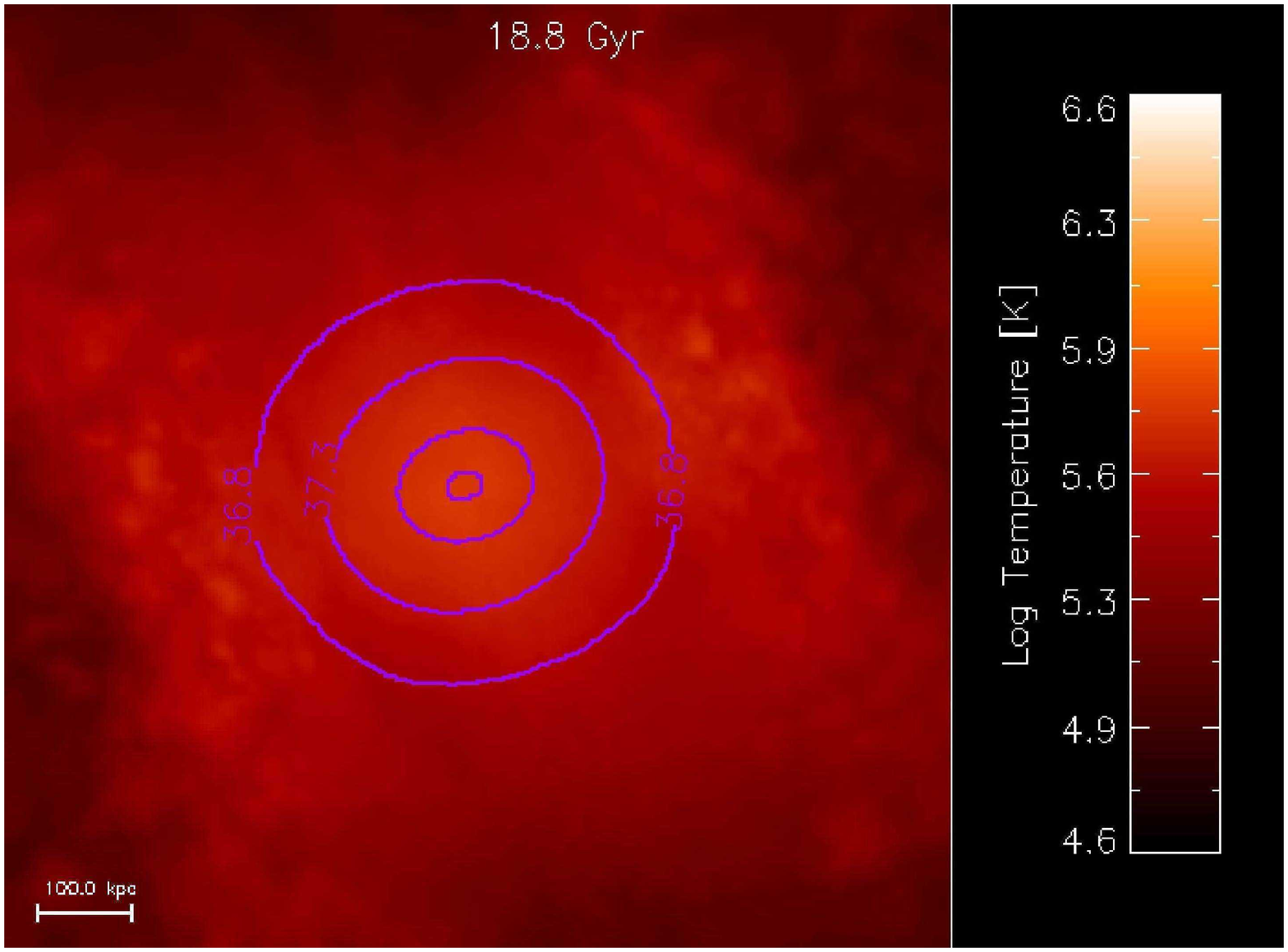}%
\\
\caption[Temperature, L$_{\rm X}$ and unbound gas for 18\% gas equal-mass merger with 2.3 kpc impact parameter.]
{\small Temperature projection for the 18\% gas fraction for a hyperbolic equal-mass merger
with an impact parameter of 2.3 kpc. This run has the highest X-ray emission 
and largest unbound material. The X-ray contours are shown in purple,the unbound material 
with white contours and the portion  of unbound material that is shocked with green contours. The shocked unbound only shows
up in the 4th image. The unbound material is identified at the end of
the simulation and then traced through the entire run. Most of the unbound material comes
from two distinct regions: the far lobes of the galaxy and the shocked front in between the galaxies. 
The projected X-ray emission also exceeds $10^{38}$ erg/s in the shocked region with a total shocked X-ray emission 
exceeding $10^{40}$ erg/s.}
\label{fig:prim18percent3quan}
\end{figure}

\begin{figure}
\centering
\includegraphics[width=3.2in,clip=true]{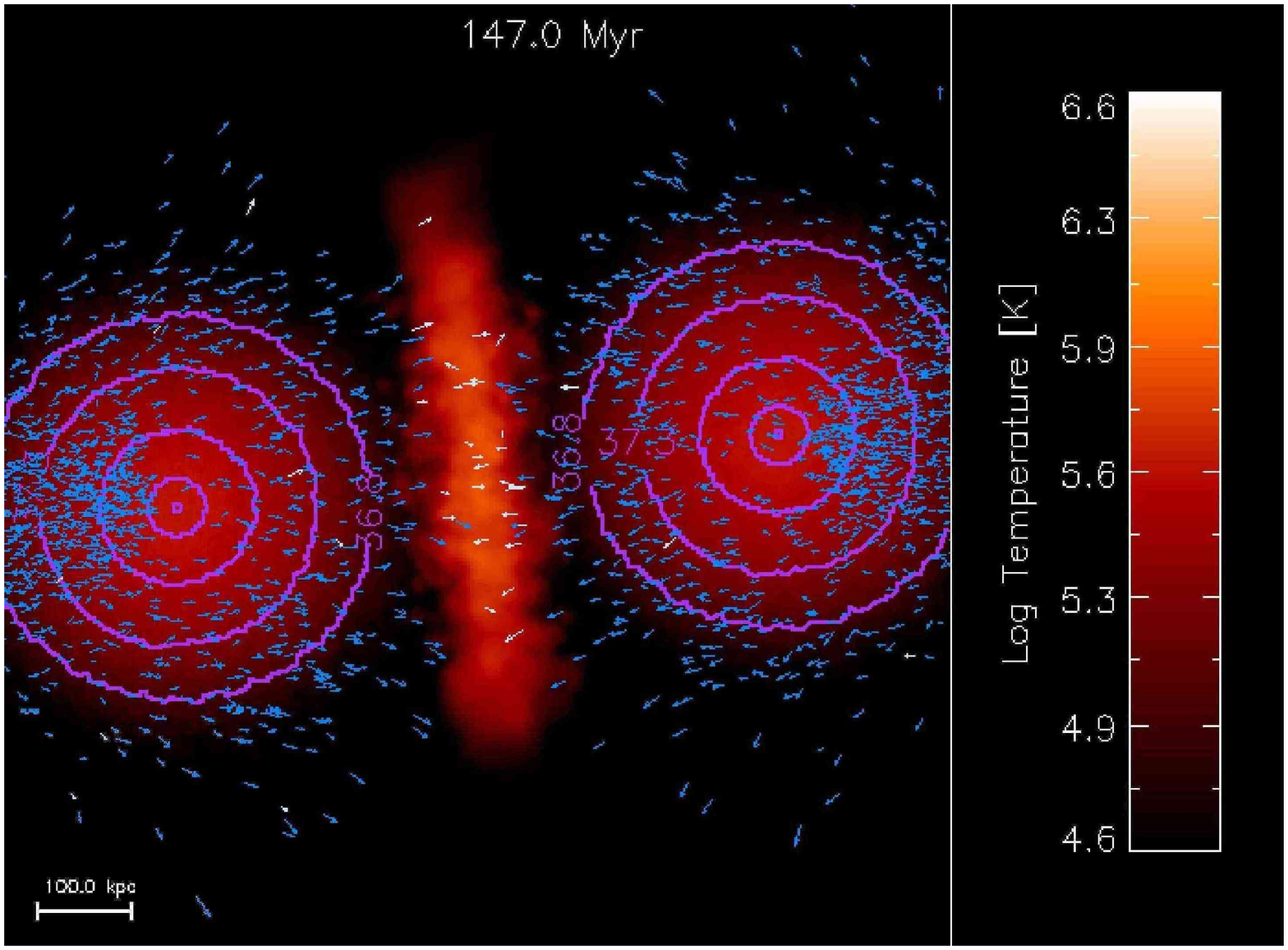}%
\includegraphics[width=3.2in,clip=true]{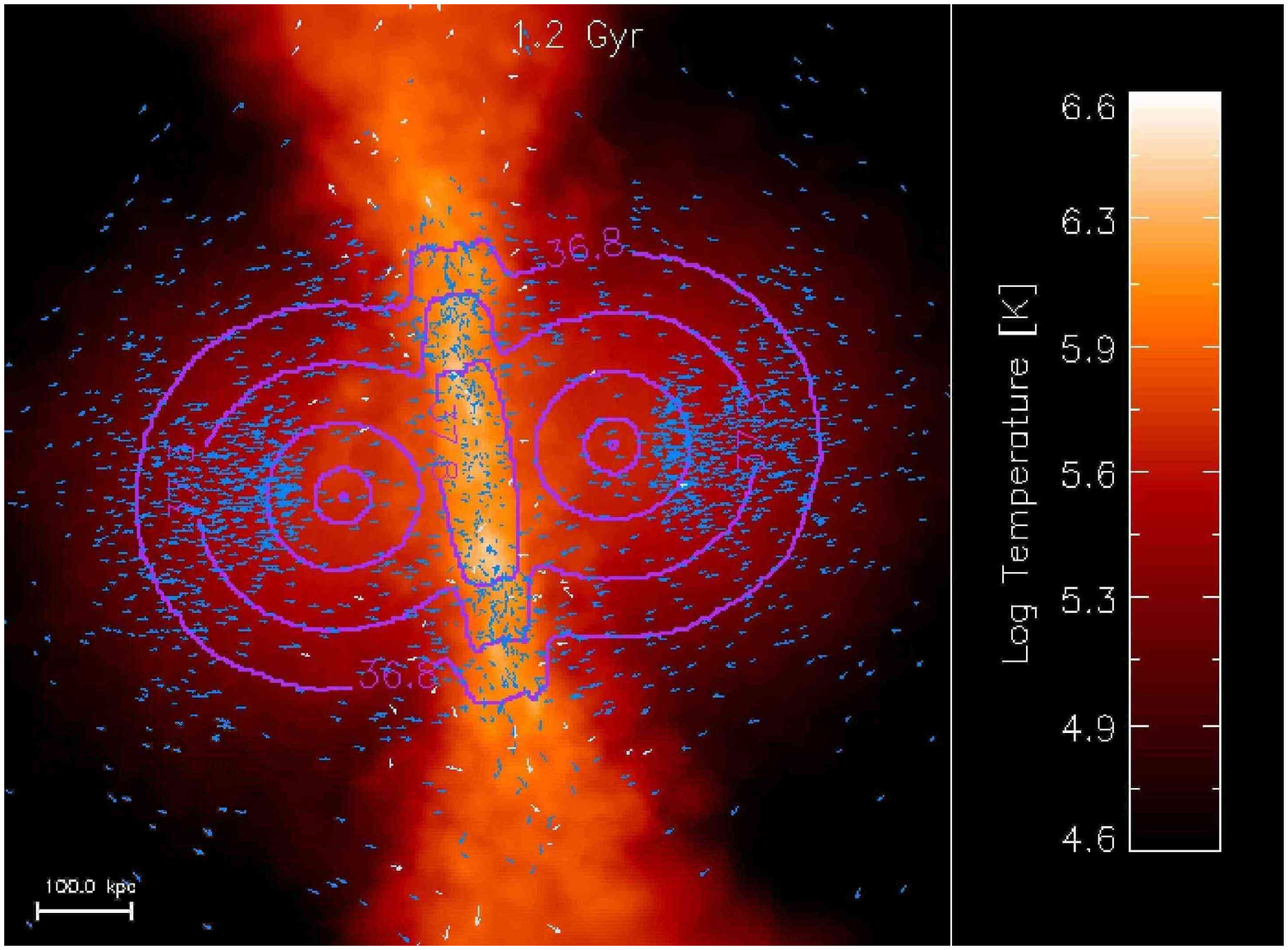}%
\\
\includegraphics[width=3.2in,clip=true]{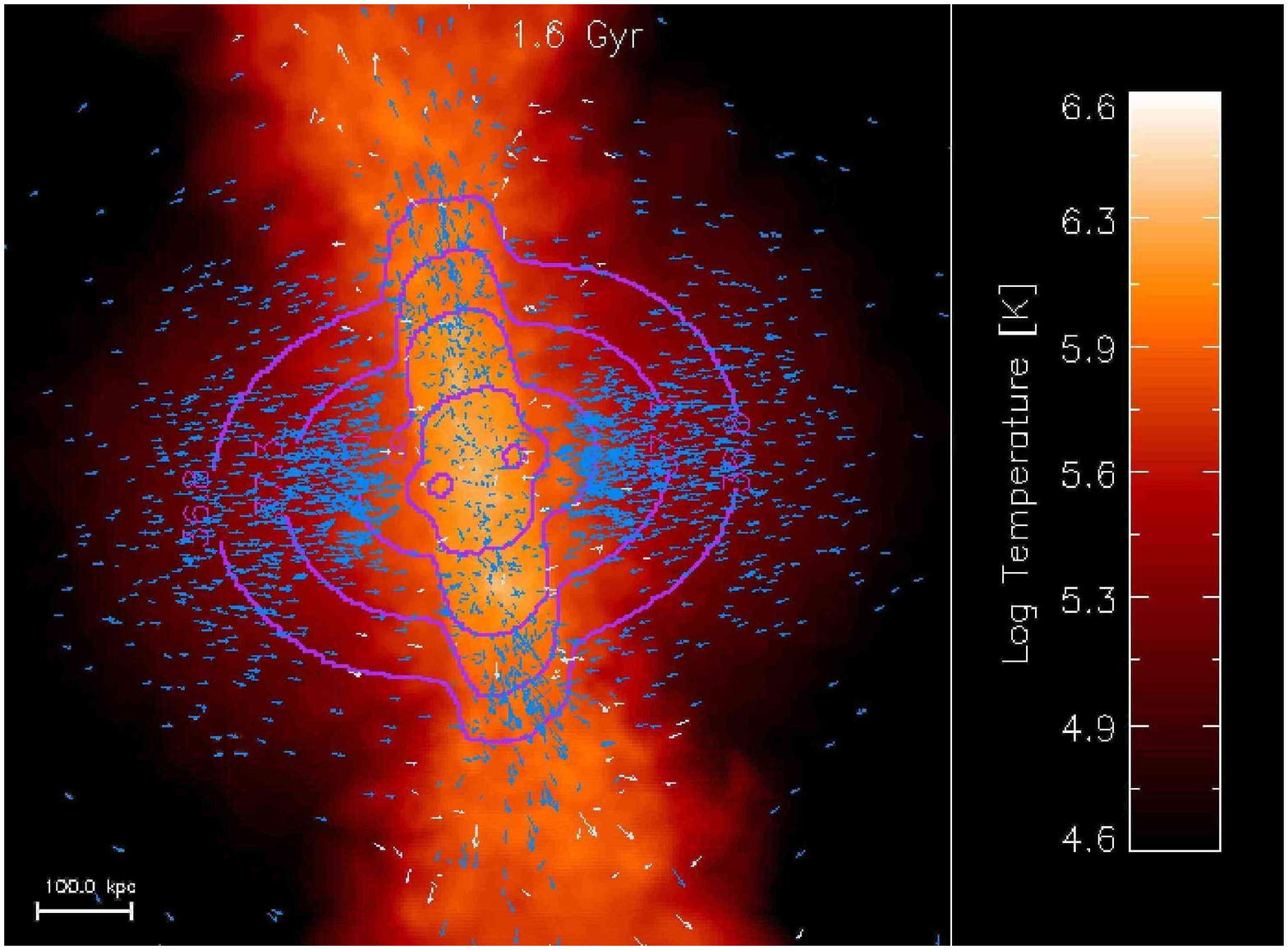}%
\includegraphics[width=3.2in,clip=true]{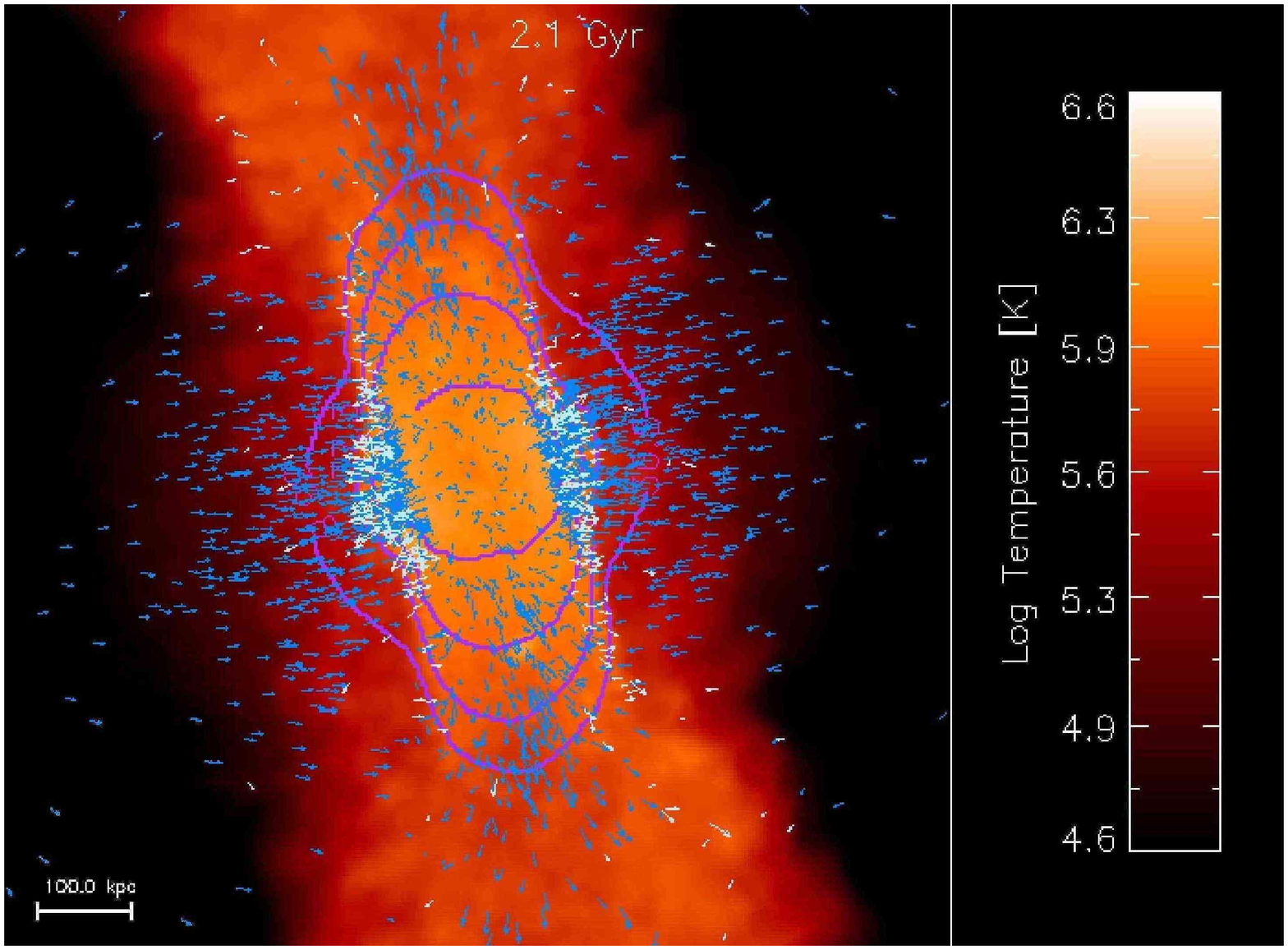}%
\\
\includegraphics[width=3.2in,clip=true]{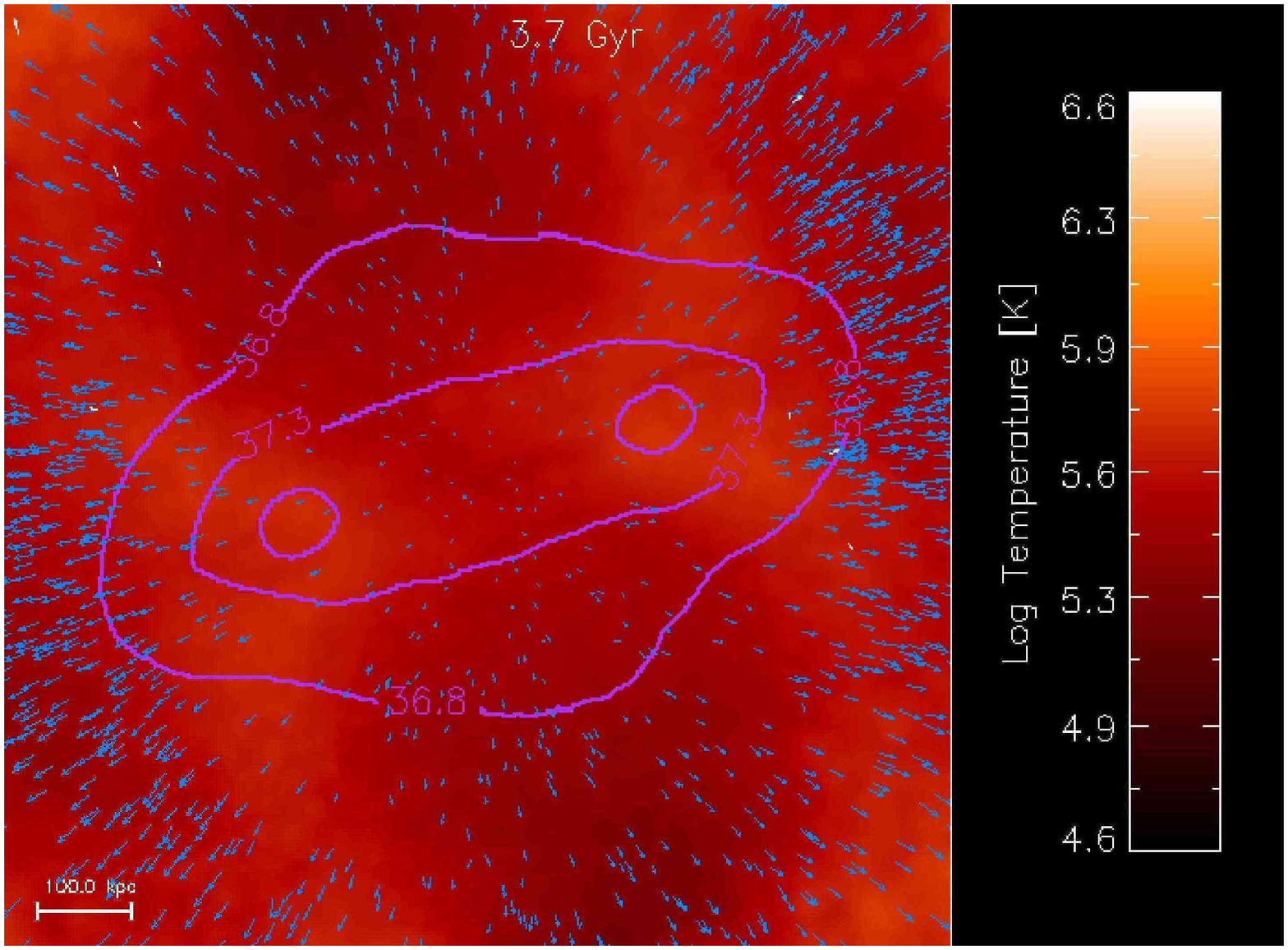}%
\includegraphics[width=3.2in,clip=true]{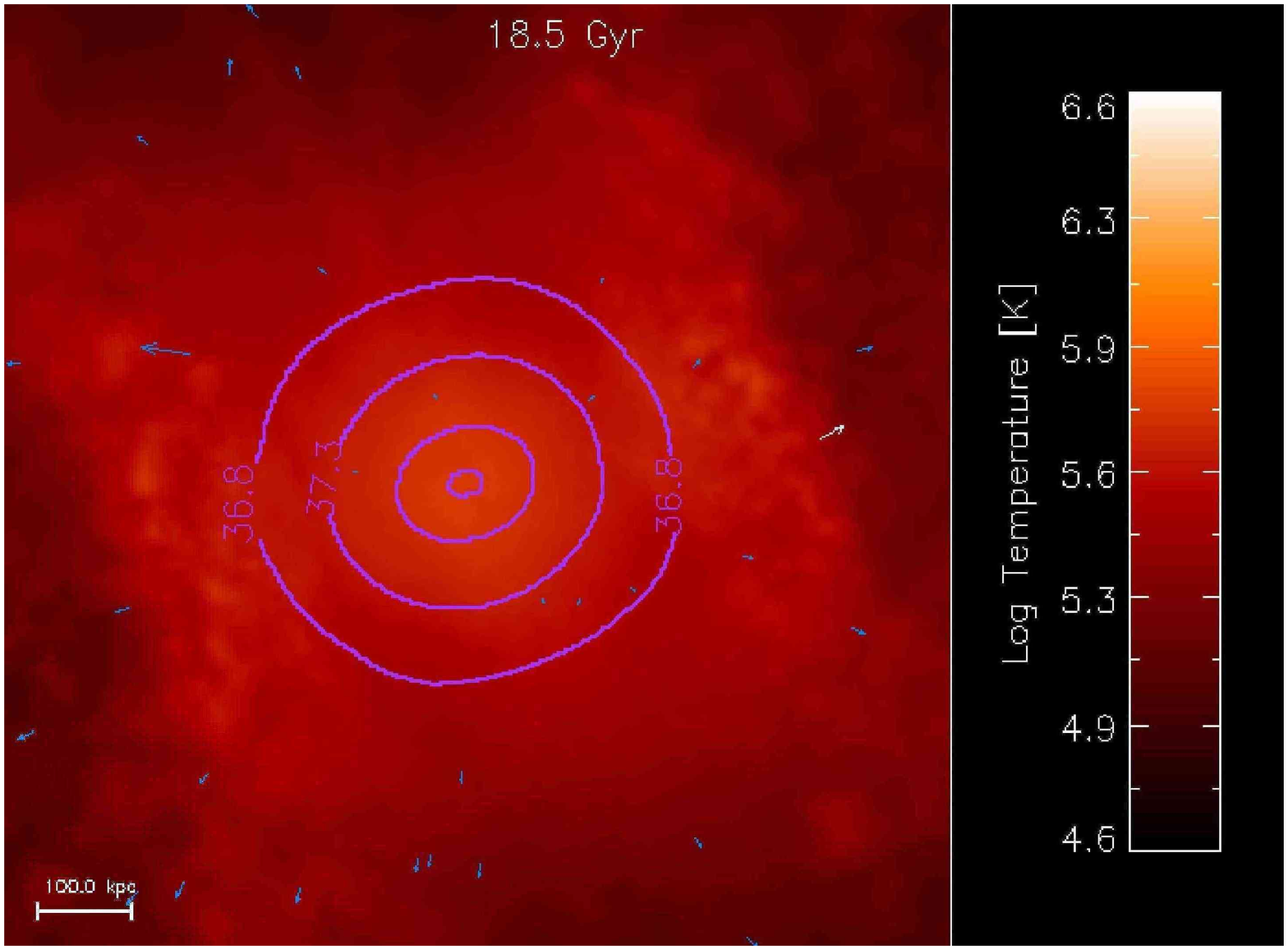}%
\\
\caption[Temperature, L$_{\rm X}$ and unbound gas with velocity vectors for 18\% gas equal-mass merger with 2.3 kpc impact parameter.]
{\small Temperature projections for the 18\% gas fraction for a hyperbolic equal-mass merger
with an impact parameter of 2.3 kpc. This run has the highest X-ray emission and largest unbound material. 
The velocities of a fraction of randomly chosen unbound particles
is plotted; blue represents unshocked unbound material while white shows the shocked unbound material. We see that
the unbound material escapes preferentially along the shocked front away from the plane of the merger since this
provides the least resistance in terms of gas pressure. Strong shocks are generated once the centres have passed and
material is ejected that interacts with the bound material at the other end of the galaxy that is yet to make a
close pass (and hence has a velocity vector opposite to that of the unbound material). This is seen in
the right image in the middle row. All of the unbound material leaves the vicinity of the galaxy much before
they actually merge ($\lesssim$ 6 Gyrs).}
\label{fig:prim18percentvel}
\end{figure}

\begin{figure}
\centering
\includegraphics[width=3.2in,clip=true]{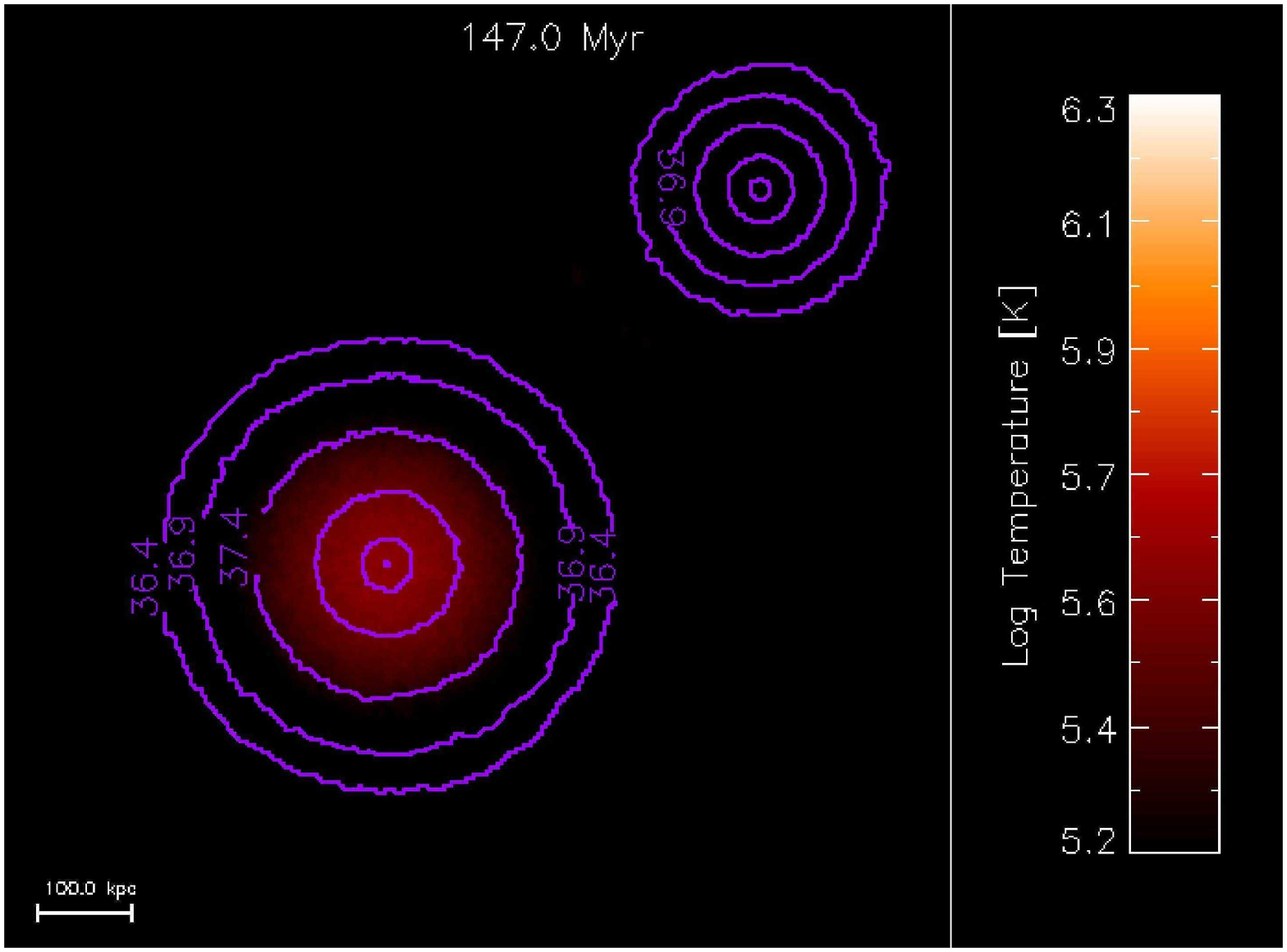}%
\includegraphics[width=3.2in,clip=true]{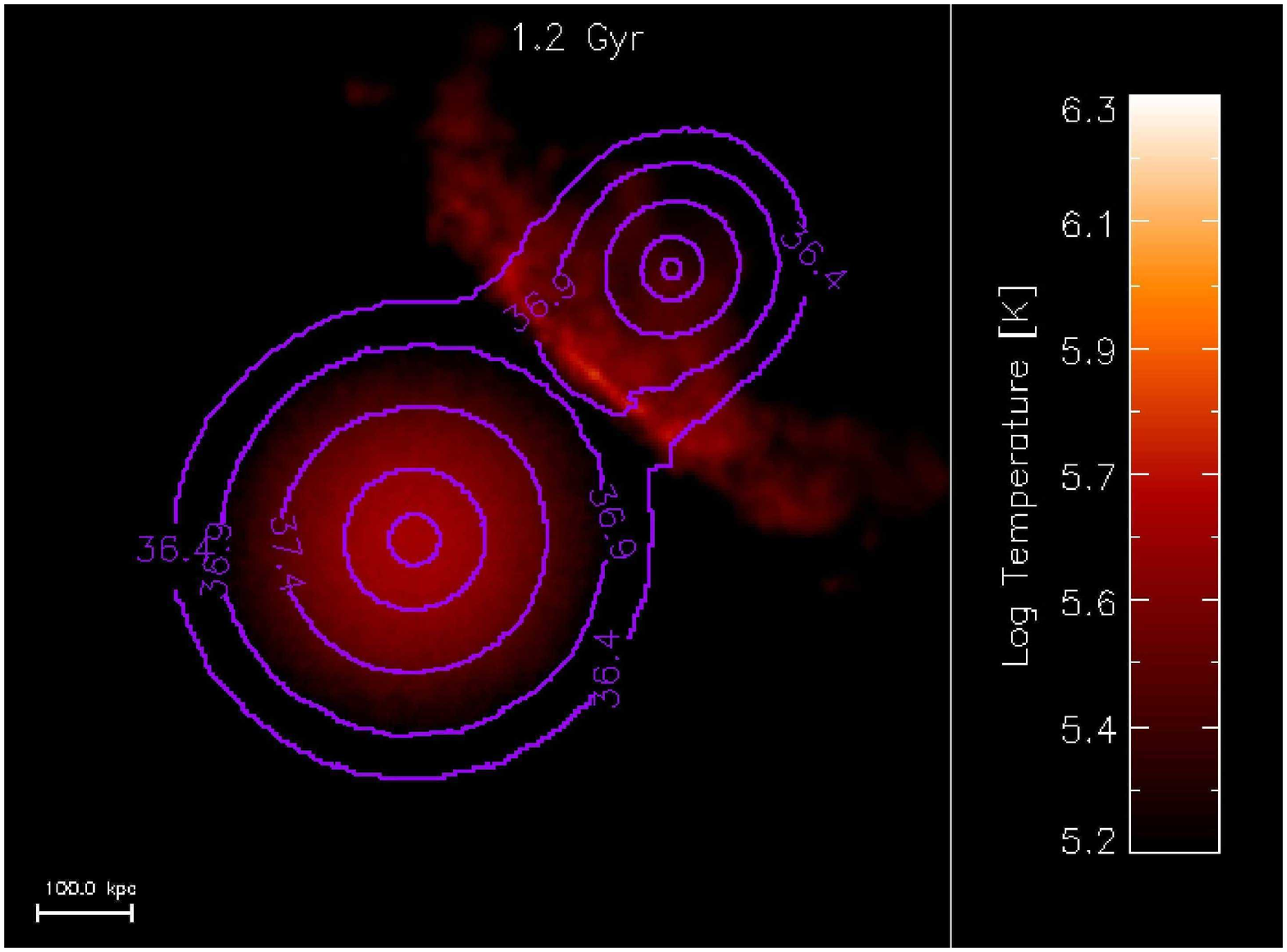}%
\\
\includegraphics[width=3.2in,clip=true]{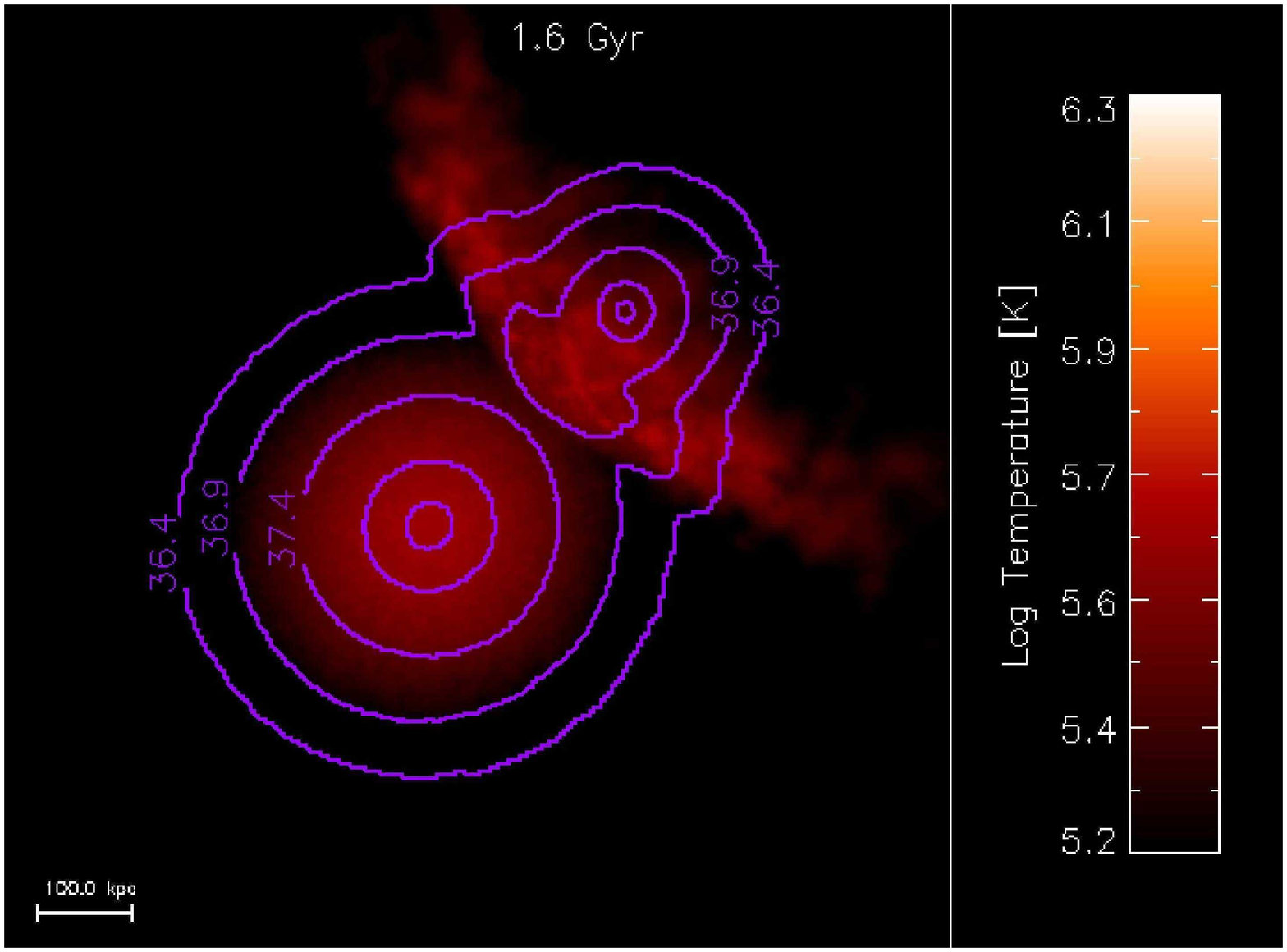}%
\includegraphics[width=3.2in,clip=true]{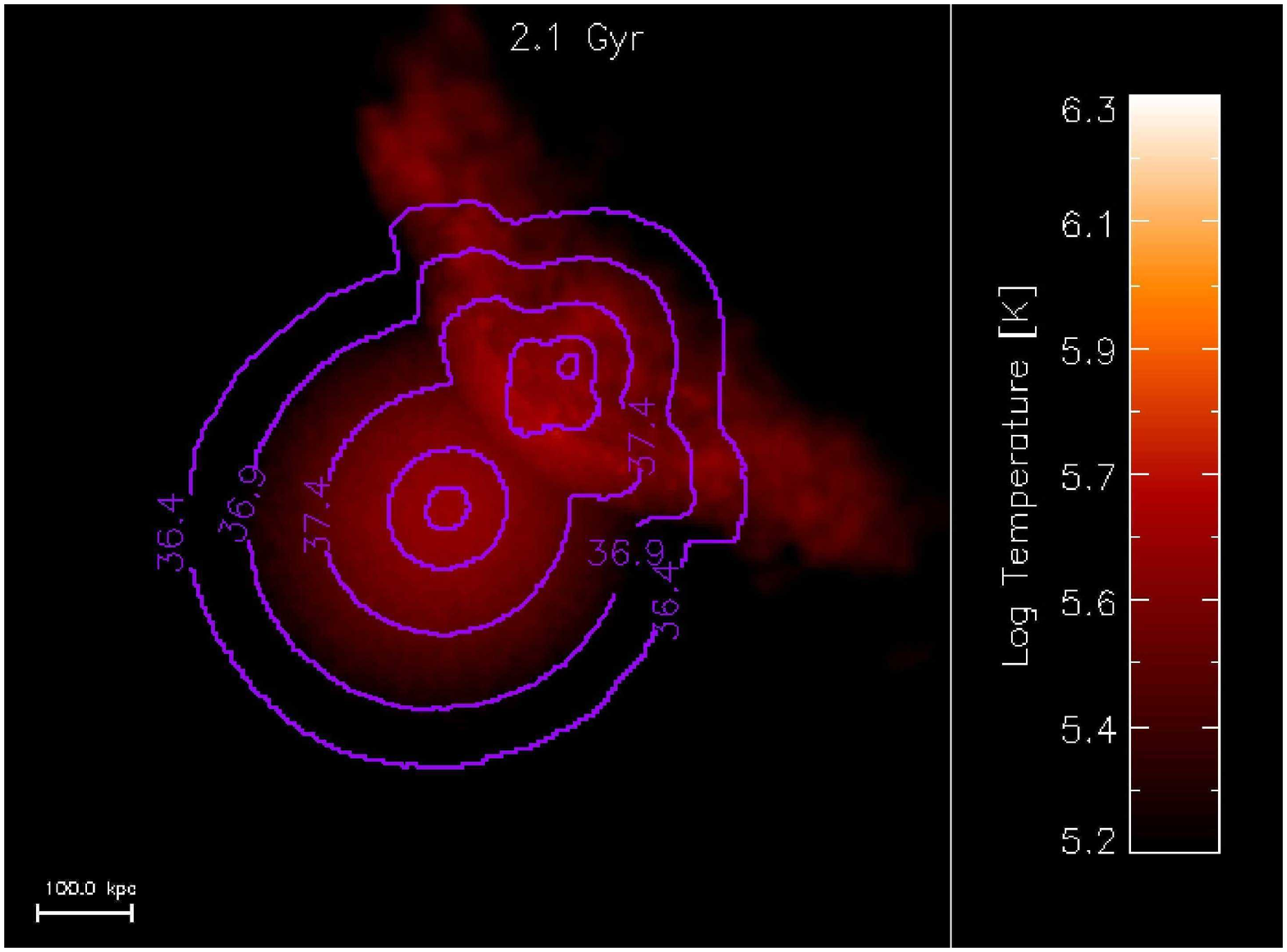}%
\\
\includegraphics[width=3.2in,clip=true]{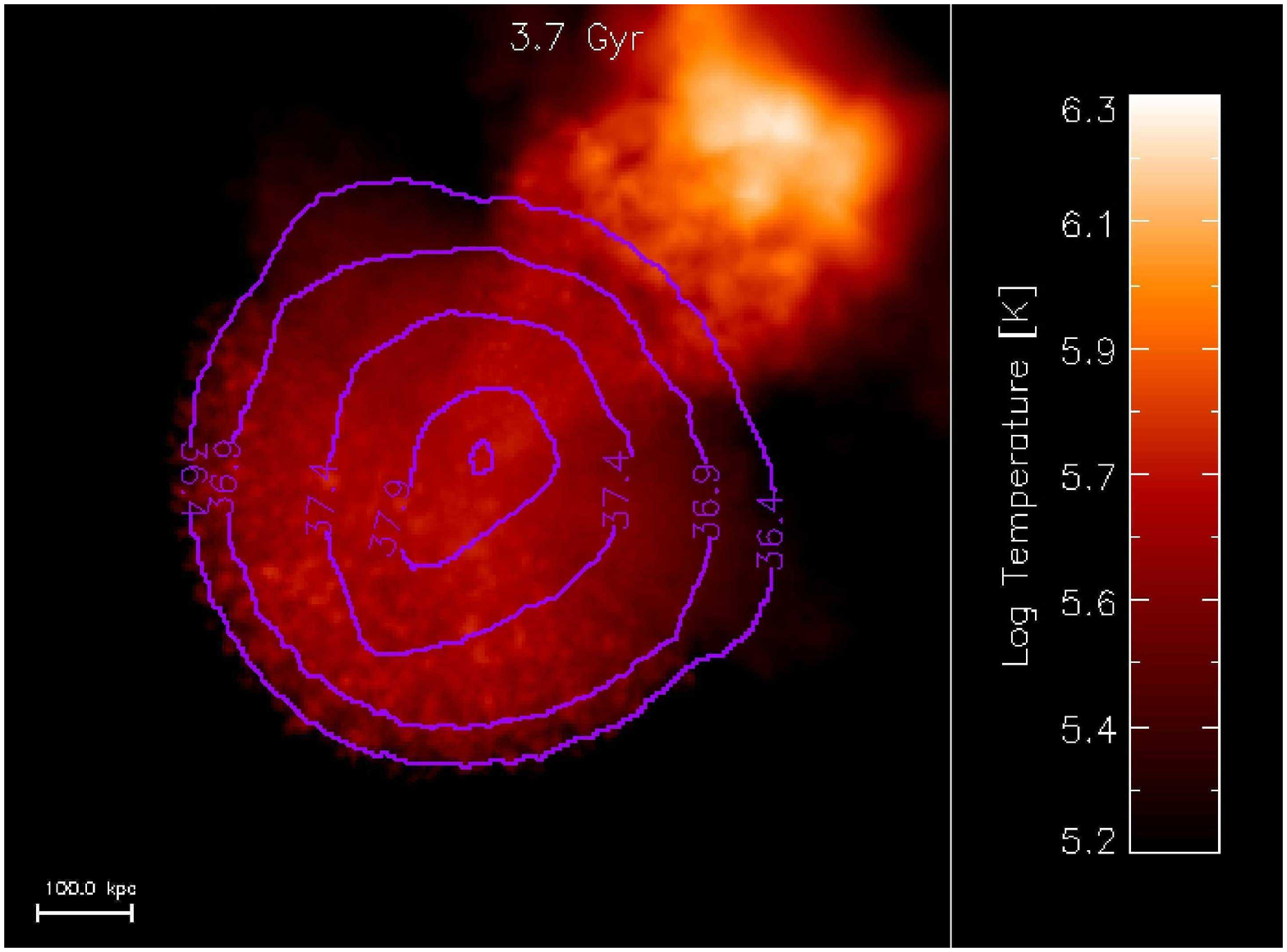}%
\includegraphics[width=3.2in,clip=true]{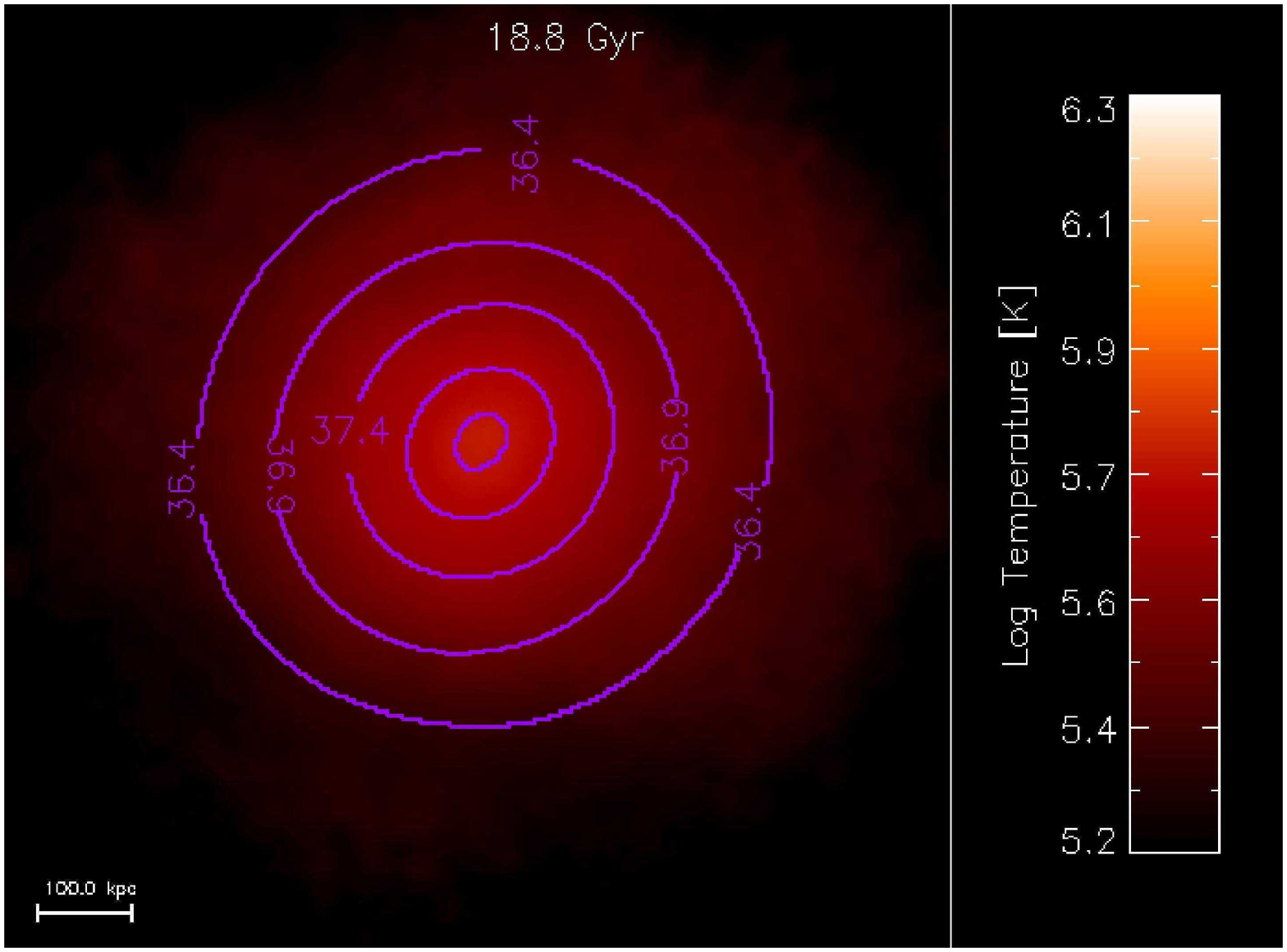}%
\\
\caption[Projected temperature and L$_{\rm X}$ contours for a 3:1 merger 18\% gas with 1.9 kpc impact parameter.]
{\small Projected temperature and L$_{\rm X}$ contours for a 3:1 merger with 18\% gas fraction and an 
impact parameter of 1.9 kpc. A prominent bow shock develops as the secondary galaxy passes through the halo of the primary. 
Even at 2.2 Gyr, when the centres are separated by $\sim$ 100 kpc, the strong shock persists and can be easily seen 
in the temperature map. Once the secondary material passes by the primary centre, a lot of material is ejected
that gets shock-heated by the secondary galaxy as it moves inwards along the orbit. This can be seen
in the patch of very bright material in the left-bottom image. The final image at 18.8 Gyr shows the merger remnant
completely relaxed with spherical L$_{\rm X}$ contours. The minimum value of the L$_{\rm X}$ contour is
 $10^{36.4}$ erg/s and increases in factors of $10^{0.5}$ erg/s.}
\label{fig:onethird18perimpzeroonetemplx}
\end{figure}

The equal-mass mergers create the strongest shocks in our entire suite of simulations. However, the surprising
part is that the strongest temperature enhancements occur after the galaxy centres have 
passed by. At this point the shocked material encounters a lower density material upfront 
and accelerates forward. The material in the far side of the galaxy is
still continuing in orbit with a velocity opposite to that of the accelerated shocked
material. When these two fluids collide, the strongest shocks (measured by the rate of entropy
change) are produced (see Fig.~\ref{fig:prim18percent3quan} and \ref{fig:prim18percentvel}). 
This can also be seen from the temperature projections in Fig.~\ref{fig:onethird18perimpzeroonetemplx}
where the highest temperatures are created by the gas that is thrown forward  after the
pericenter passage, colliding with the gas in the far side of the galaxy. 
Consequently, the peak $\LX$ from shocks occurs some time after the peak $\LX$. 
The equal-mass mergers with 1\%, 10\% and 18\%  gas have a peak $\LX$ 
from shocks of $\sim 10^{37}$, $\sim 10^{39}$ and $\sim 3\times10^{39}$ erg/s. 

The 3:1 mergers show a similar pattern; with $\LX$ increasing with decreasing impact parameter and
increasing gas fraction. The peak $\LX$ for the 3:1 mergers with 1\% gas is $\sim 10^{39}$ erg/s
while the peak $\LX$ from shocks is $\sim 5\times 10^{36}$ erg/s. The 10\% gas fraction
produces a peak $\LX$ of $\sim 10^{41}$ erg/s and $\sim 10^{38}$ is from shocks. For the 18\%
gas fraction, the peak $\LX$ and the peak $\LX$ from shocks are $\sim 10^{42}$ and $\sim 10^{39}$
erg/s respectively. Though peak $\LX$ from shocks is larger than $10^{39}$ erg/s it is 
still 3 orders of magnitude smaller than the total $\LX$ signature from the merger event. 

The 10:1 mergers follow the trend and for the 1\% gas fraction simulations the peak $\LX$ and the peak
$\LX$ from shocks are $\sim 2.0\times 10^{39}$ and $\sim 10^{36}$ erg/s respectively. For the
10\% simulations the peak $\LX$ is $\sim 2 \times 10^{41}$ erg/s while the peak $\LX$ from shocks is $\sim 3\times10^{37}$ erg/s. 
For the 18\% gas fractions, the peak $\LX$ is $\sim 5 \times10^{41}$
erg/s while the peak $\LX$ from shocks is $\sim 2\times 10^{38}$ erg/s. 
Clearly, the 10:1 mergers do not produce any appreciable
X-ray signature from the shock-heating of the hot gas even with the largest gas fractions. See Table~\ref{table:mergers}
for detailed data on X-ray production for all the simulations.

From our simulations, we see that the expected trends do occur. The X-ray luminosity, both intrinsic
and from shocks, increases
with increasing halo mass and decreasing impact parameters. Larger initial velocities do
generate larger $\LX$; however, if the hot gas fraction drops too much, as it does in the case of some
of the hyperbolic mergers, the X-ray emission can be reduced. In light of these trends, the best bet 
to detect the shock signature of hot halo gas during mergers would be to locate major mergers
between Milky Way-type galaxies that are happening in the local universe. Such a system would produce significant
X-rays from the hot halo and the shock-heated interface if the galaxies are to be observed in the initial
stages of the merger. Spatial resolution will not be an issue since the signature is an extended source; the
problem would be the identification of the underlying shock X-ray emission and separate it from the
emission from the hot halo particles themselves. {\it XMM-Newton}, with its superior collecting area, would
have been the instrument of choice; however, the inferior sensitivity of {\it XMM-Newton}  in the soft
X-ray band, where most of the shocked gas in our simulation radiates (the maximum temperature reached
in our simulations is $\sim 0.3$ keV), {\it Chandra} is best suited 
to characterise the hot halo gas. Our threshold for the X-ray flux, $10^{-15}$ erg/s/cm$^{2}$ is 
relevant to {\it Chandra}.

So far we have discussed the production of X-rays by the merger, both from the hot gas in the
halo and the shocks. We found that the X-ray emission  depends on a variety of parameters
like the merger ratio, the cumulative mass, the merger orbit and the gas fraction in the simulation.
We attempt to combine the effects of all these factors to predict the peak X-ray luminosity from 
shocks as a function of the gravitational impulse from  the merger.  \citet{CD08} derive an empirical fitting formula for
the impulse motivated by numerical simulations of galaxy mergers.
The impulse, $\Delta E$, is approximated by:
\begin{equation}
\Delta E  \propto \dfrac{G^2 M_1\,M_2^2}{v_{\rm p}^2\,\left[r_{\rm p}^2 + r_{\rm p}\times {\rm R_{vir,1}}\right]},
\label{eqn:shockimpulse}
\end{equation}
where $M_1$ is the primary galaxy mass, $M_2$ is the mass of the secondary, $v_{\rm p}$ and $r_{\rm p}$ are the
relative velocity of the galaxies and the pericenter distance as determined from the simulation. 

\begin{figure}
\centering
\includegraphics[scale=0.55,clip=true,bb=5 30 737 530]{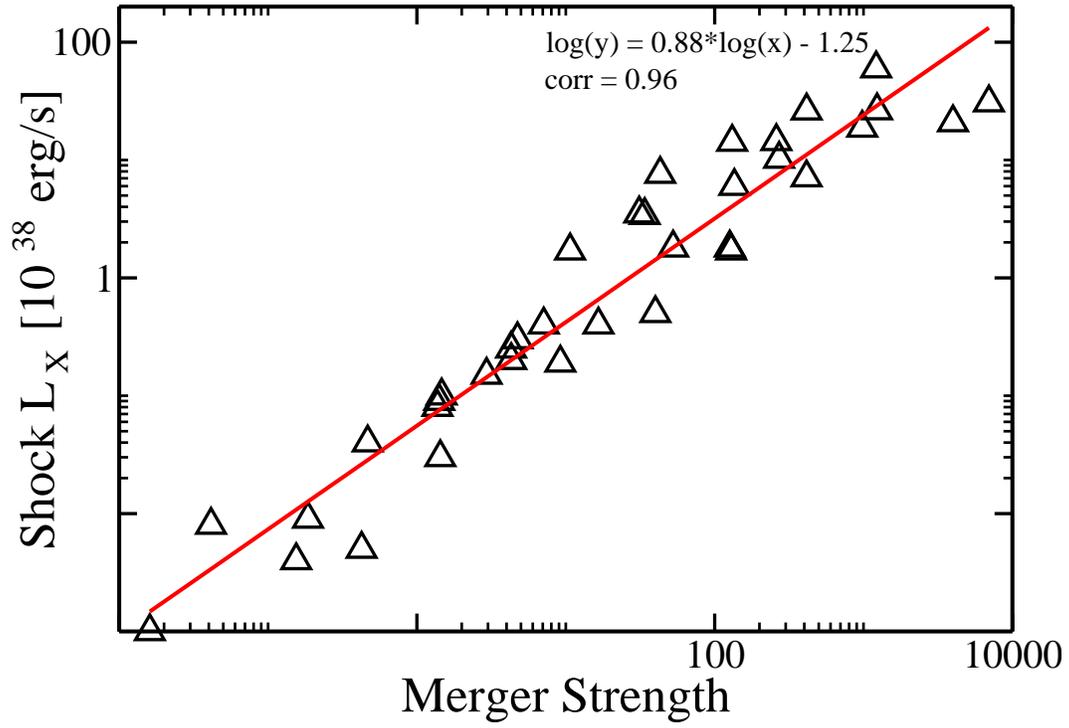}
\caption[The semi-analytic fit to the peak shock luminosity]
{\small The fit for the peak $\LX$ from shocks as a function of the
the merger strength, $\Delta H$ (see Eqn.~\ref{eqn:allshocksemifit}). Each simulation is plotted as a triangle
while the fit (see Eqn.~\ref{eqn:allshocksemifitform}) is plotted with a red line. The correlation
for this fit is 0.96 and implies that Eqn.~\ref{eqn:allshocksemifitform} is a good linear fit.}
\label{figure:allshocksemifit}
\end{figure}

We also define a ``merger strength'', H, that is logically related to $\LX$. 
$\LX$ depends on $\rho^2 \sqrt{T}$, and $T \propto v_{\rm relative}^2$ (see Eqn.~\ref{eqn:rankinetemp}). 
Therefore, $\LX$ should depend on $v_{\rm relative}$; we use the velocity at perihelion, $v_{\rm p}$ determined from the simulation
to capture this. Our final equation for the merger strength, $\Delta H$, is then $\Delta E \, f_{\rm gas}^2 \,v_{\rm p}$,
and can be written as:
\begin{equation}
\Delta H \propto  \dfrac{G^2 M_1\,M_2^2\,f_{\rm gas}^2}{v_{\rm p} \,\left[r_{\rm p}^2 + r_{\rm p}\times {\rm R_{vir,1}}\right]},
\label{eqn:allshocksemifit}
\end{equation}
We then fit the peak $\LX$ from shocks, $L_{\rm peak}$, as a linear function of $\Delta H$. However, a better
fit, with a correlation co-efficient of 0.96,  is obtained with a logarithmic fit. The fit for our set of 36 simulations is:
\begin{equation}
\log_{10}\,\left[\frac{L_{\rm peak}}{10^{38} \;{\rm erg/s}}\right] = 0.88 \times \log_{10}\,\Delta H - 1.25 .
\label{eqn:allshocksemifitform}
\end{equation}
This fit captures essential  physics  in a gas-rich galaxy merger and can be used  
to estimate the peak $\LX$ from shocks in a merger (see Fig.~\ref{figure:allshocksemifit}). Such a fit is quite useful in modelling the total
$\LX$ from shocks in semi-analytic studies of galaxy formation. 

\subsection{Unbound Gas}\label{section:unbndgas}

To constrain the amount of unbound material resulting from the mergers, we evolve the
corresponding initial conditions in isolation for a similar time-scale. Every galaxy will
lose some small amount of material from spurious two-body relaxation, and by evolving the galaxies in
isolation, we can estimate the magnitude of this effect. We deduct this
amount of unbound material (column 3 in Table~\ref{table:isolated2}) when we  compute the final unbound fraction
from the merger (see column 4 in Table~\ref{table:shocked}).

\begin{center}
\renewcommand{\thefootnote}{\fnsymbol{footnote}}
\scriptsize
\begin{longtable}{cccccccc}
\caption[Final stage of isolated galaxies after 15 Gyr]{
Isolated galaxy evolution.
We evolve isolated galaxies with AVP$=0.5$ for $\gtrsim$ 10 Gyr to ascertain
the amount of unbound material from internal processes. 
This fraction is subtracted from the unbound material from
the mergers, we can fix the actual amount freed during the merger itself.}
\label{table:isolated2} \\

\hline \hline \\[-2ex]
   \multicolumn{1}{c}{\textbf{Gas Content}} &
   \multicolumn{1}{c}{\textbf{Galaxy Type}} &
   \multicolumn{1}{c}{\textbf{Unbound gas }} &
   \multicolumn{1}{c}{\textbf{L$\mathbf{_{\rm X}}$}} &
   \multicolumn{1}{c}{\textbf{R$\mathbf{_{\rm vir}}$}} &
   \multicolumn{1}{c}{\textbf{M$\mathbf{_{\rm vir}}$}} &
   \multicolumn{1}{c}{\textbf{Hot gas}} \\[0.5ex] 
   \multicolumn{1}{c}{\textbf{--}} &
   \multicolumn{1}{c}{\textbf{--}} &
   \multicolumn{1}{c}{\textbf{[\%]}} &
   \multicolumn{1}{c}{\textbf{[$\mathbf{10^{40}}$ erg/s]}} &
   \multicolumn{1}{c}{\textbf{[kpc]}} &
   \multicolumn{1}{c}{$\mathbf{ [10^{10} \,\Msun ]}$} &
   \multicolumn{1}{c}{\textbf{[\%]}} \\[0.5ex]\hline \hline \\[-2ex]
\endfirsthead

\multirow{3}{*}{1\% gas}
& Primary  \footnotemark[2]   & 0.43 & 0.09   & 214.8 & 113.2  &  93.2   \\
& Onethird                    & 0.69 & 0.03   & 149.1 & 37.8   &  73.4   \\
& Onetenth                    & 0.84 & 0.01   & 100.2 & 11.5   &  41.3   \\[0.7ex] \hline 
\multirow{3}{*}{10\% gas} 
& Primary   \footnotemark[3] & 0.60 & 8.46   & 212.9 & 110.2  &  89.8    \\
& Onethird                   & 0.61 & 3.38   & 149.1 & 37.7   &  75.9    \\
& Onetenth                   & 0.83 & 0.94   & 100.7 & 11.7   &  45.3    \\[0.7ex] \hline 
\multirow{3}{*}{18\% gas}
& Primary                    & 0.72 & 27.5   & 213.9 & 112.0  &  95.2    \\
& Onethird                   & 0.60 & 11.3   & 148.6 & 37.6   &  79.0    \\
& Onetenth                   & 0.76 & 3.05   & 100.7 & 11.6   &  48.4    \\[0.7ex]
\hline \hline \\[-2ex]
\footnotetext[2]{at 15 Gyrs.}
\footnotetext[3]{at 10 Gyrs.}
\end{longtable}
\normalsize
\renewcommand{\thefootnote}{\arabic{footnote}}
\end{center}

\begin{figure}
\centering
\includegraphics[scale=0.55,clip=true,bb= 0 0 743 572]{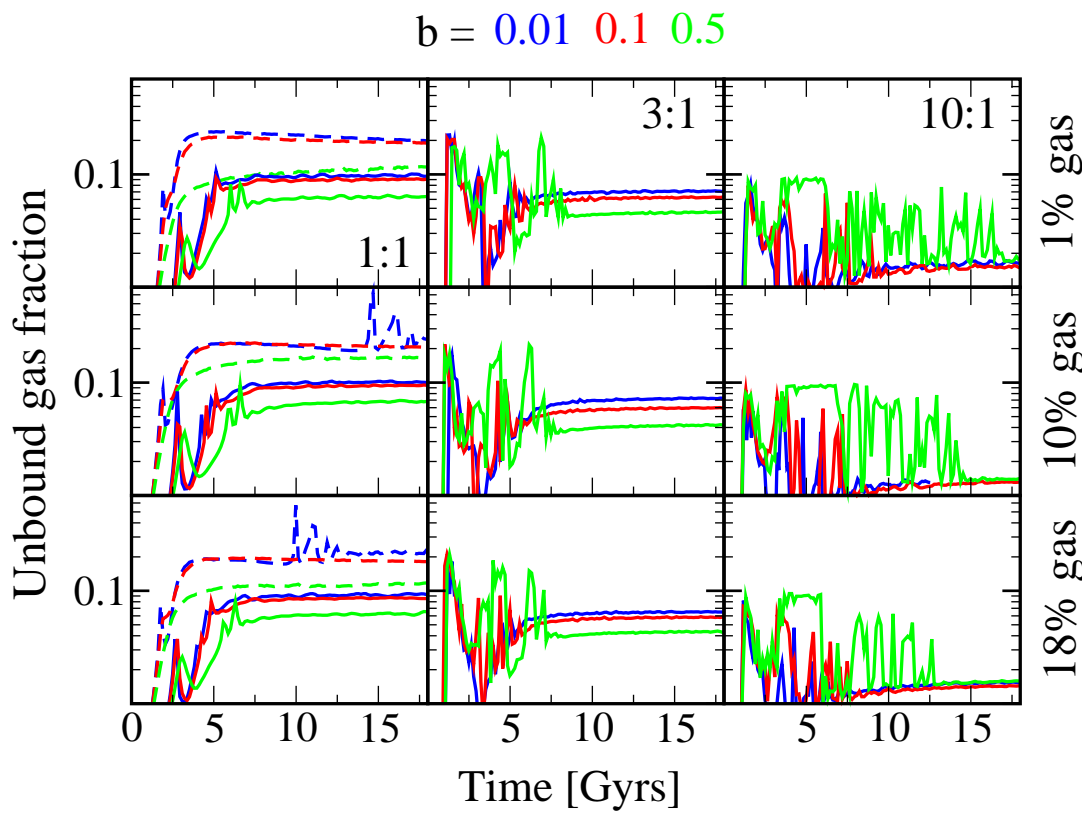}
\caption[Evolution of unbound gas fraction for all simulations]
{\small Evolution of the unbound gas fraction for all simulations. Equal-mass
mergers with the 0.01 \& 0.1 $\Rvir$ has $\sim$ 20\% unbound
gas fraction from the final merger remnant. The 3:1 mergers release
about 5-7\% gas while the 10:1 mergers unbind the least amount
of gas at about 0.5\%. The irregular nature of the unbound gas
fraction for the 3:1 and the 10:1 mergers is caused by the
multiple passages taken by the secondary galaxy.}
\label{fig:allsimsunboundgas}
\end{figure}

We find that most of the unbound material originates at or near the virial radius of the respective 
galaxies. As the merger progresses, this material escapes to
beyond $3\,R_{\rm vir}$ even before the second passage. 
By the end of the simulation, this unbound gas is far ($\gtrsim 1$ Mpc) from the remnant 
(see Fig.~\ref{fig:prim18percent3quan}), 
thereby alleviating the problem of transporting gas to inter-galactic distances \citep{FPS00}. 
We find that $\sim \,9\%$ of the gas unbinds, along with a similar fraction in the dark matter. 
Per merger, this translates to a gas mass  $\gtrsim 3\,\times 10^{10}$\,$\Msun$ from equal-mass
and $\gtrsim 10^{10}$ $\Msun$ from 3:1 mergers flowing into the IGM (see Fig.~\ref{fig:allsimsunboundgas}). While at 18 Gyrs our simulation
lasts much longer than a Hubble time, nearly all the unbound material flows outside $3\,R_{\rm vir}$
within the first 5-6 Gyrs. Considering that every galaxy has had a major merger in the past \citep[see]{CBDP03},
this may be an excellent mechanism to enrich the IGM.
Thus galaxy mergers play an important role not only in the evolution of the galaxies and  hierarchical
structure formation, but they may also influence both the mass and the metallicity content of the IGM.

We find that the unbound material originates from two distinct regions -- from the near and
far lobes of the galaxy along their line of motion. These constitute the regions that are shocked
during the course of the merger. The near lobes are shocked before pericenter passage, which can be
seen in Fig.~\ref{fig:prim18percent3quan}. These shocked gas particles then leave in directions
perpendicular and away from the direction of the motion of the galaxies as they expand into the
lower density regions. Their escape is facilitated by the lower density and pressure of the ambient
medium above the shocked region and can be seen from the velocity vectors of the unbound particles
pointing outwards in Fig.~\ref{fig:prim18percentvel}. The gas particles from this region
escape from the galaxies the earliest.

To understand why the gas escapes from the far regions of the galaxies,
we must look into the shock process itself. As the galaxies pass through pericenter,
the density in front of the shocked material suddenly drops rapidly. This causes the shocked
material to accelerate, and it is flung forward. However, the material at the
far side of the galaxy is still moving with the bulk flow. This naturally results in the
largest relative velocity and hence, the strongest shocks in the simulation. The original shocked
gas interacts with the gas from the far side, and the gas from the far side receives a large
impulse, which causes the gas to flow outward instead of continuing with the bulk motion of its host galaxy. 
This material can be seen as the brightest patches in the temperature images (see Fig.~\ref{fig:prim18percentvel} 
and ~\ref{fig:onethird18perimpzeroonetemplx}).
In unequal-mass  mergers, the impulse affects the lower mass galaxies more, and an asymmetry
is introduced whereby the strongly-shocked unbound material comes primarily
from the smaller galaxy. This can be seen in the images of the temperature projections for
the unequal-mass mergers where the brightest patches are towards the side of the smaller
galaxies (see Fig.~\ref{fig:onethird18perimpzeroonetemplx} for the 3:1 case). 
Although most of the unbound material originates from a radius greater than 0.5 $\Rvir$ in each
galaxy and therefore should have a systematically lower metallicity, the gas can get enriched
during the merger process itself from starburst-driven winds.  

\clearpage
\newpage

\begin{center}
\renewcommand{\thefootnote}{\fnsymbol{footnote}}
\renewcommand{\arraystretch}{0.79}
\scriptsize
\begin{longtable}{ccccccc}
\caption[Table for all the simulations showing the fraction of shocked unbound gas]
{Unbound gas from mergers.  The first three equal-mass mergers for each gas fraction are the hyperbolic
encounters while the rest are all bound elliptical orbits with a fixed amount
of energy in the orbit for a given merger ratio. The unbound fraction (Column 4)
is obtained by deducting the unbound gas fractions when evolved in isolated 
(see Table~\ref{table:isolated2}) from the
fraction determined at the end of the simulation. Column 5 shows the fraction
of all the gas particles that are shocked during the simulation.
The unbound gas particles are tracked through the snapshots to find where
they become unbound or shocked. The fraction of unbound gas particles
that are shocked during the simulation is shown in Column 6.}
\label{table:shocked} \\

\hline \hline \\[-2ex]
   \multicolumn{1}{c}{\textbf{Gas}} &
   \multicolumn{1}{c}{\textbf{Merger Type}} &
   \multicolumn{1}{c}{\textbf{b}} &
   \multicolumn{1}{c}{\textbf{Unb. frac.}} &
   \multicolumn{1}{c}{\textbf{Shocked (all)}} &
   \multicolumn{1}{c}{\textbf{Shocked (unb)}} \\[0.8ex]
   \multicolumn{1}{c}{\textbf{--}} &
   \multicolumn{1}{c}{\textbf{--}} &
   \multicolumn{1}{c}{\textbf{[kpc]}} &
   \multicolumn{1}{c}{\textbf{[\%]}} &
   \multicolumn{1}{c}{\textbf{[\%]}} &
   \multicolumn{1}{c}{\textbf{[\%]}} \\[0.5ex]\hline \hline \\[-2ex]
\endfirsthead
\multirow{12}{*}{\begin{sideways}{\Large{1\% gas}} \end{sideways}}
& 1:1h  & 2.3   &      19.2 &     13.1    &     49.6  \\
& 1:1h  & 22.8  &      18.5 &     11.1    &     46.7  \\
& 1:1h  & 114.3 &      11.2 &      5.9    &     44.1  \\
& 1:1   & 2.3   &       9.4 &     14.1    &     84.2  \\
& 1:1   & 22.8  &       8.4 &     13.8    &     85.8  \\
& 1:1   & 114.3 &       5.9 &     11.9    &     89.5  \\
& 3:1   & 1.9   &       6.7 &      9.4    &     69.6  \\
& 3:1   & 19.4  &       5.7 &      9.2    &     73.1  \\
& 3:1   & 96.8  &       4.1 &      8.0    &     80.0  \\
& 10:1  & 1.7   &       1.2 &      2.6    &     60.9  \\
& 10:1  & 16.7  &       1.1 &      2.5    &     62.6  \\
& 10:1  & 83.7  &       1.2 &      4.0    &     66.0  \\[0.1ex]\hline\\[0.1ex]
\multirow{12}{*}{\begin{sideways}{\Large{10\% gas}} \end{sideways}}
& 1:1h   & 2.3   &      23.8 &     22.5   &     53.6   \\
& 1:1h   & 22.8  &      20.0 &     13.4   &     53.4   \\
& 1:1h   & 114.3 &      16.2 &      8.4   &     44.8   \\
& 1:1    & 2.3   &       9.7 &     14.6   &     86.9   \\
& 1:1    & 22.8  &       8.9 &     14.1   &     86.8   \\
& 1:1    & 114.3 &       6.4 &     12.1   &     87.7   \\
& 3:1    & 1.9   &       6.8 &      9.1   &     67.2   \\
& 3:1    & 19.4  &       5.5 &      7.2   &     63.6   \\
& 3:1    & 96.8  &       3.7 &      6.0   &     68.5   \\
& 10:1   & 1.7   &       0.8 &      1.4   &     41.0   \\
& 10:1   & 16.7  &       0.9 &      1.2   &     36.4   \\
& 10:1   & 83.7  &       1.0 &      2.0   &     39.6   \\[0.1ex]\hline\\[0.1ex]
\multirow{12}{*}{\begin{sideways}{\Large{18\% gas}} \end{sideways}}
& 1:1h   & 2.3   &      21.8 &     19.8   &     50.7  \\*
& 1:1h   & 22.8  &      17.5 &     12.2   &     52.5  \\*
& 1:1h   & 114.3 &      11.1 &      7.0   &     50.8  \\*
& 1:1    & 2.3   &       8.6 &     13.7   &     85.0  \\*
& 1:1    & 22.8  &       8.1 &     13.6   &     85.6  \\*
& 1:1    & 114.3 &       5.6 &     11.6   &     88.9  \\*
& 3:1    & 1.9   &       6.0 &      8.2   &     67.8  \\*
& 3:1    & 19.4  &       5.3 &      7.8   &     70.2  \\*
& 3:1    & 96.8  &       3.9 &      6.7   &     77.5  \\*
& 10:1   & 1.7   &       1.1 &      2.0   &     44.7  \\*
& 10:1   & 16.7  &       1.0 &      1.9   &     46.2  \\*
& 10:1   & 83.7  &       1.1 &      2.8   &     54.4  \\*[0.3ex]\hline \hline \\[-2ex]
\end{longtable}
\normalsize
\renewcommand{\thefootnote}{\arabic{footnote}}
\renewcommand{\arraystretch}{1.0}
\end{center}

It is clear that the unbound material is related to the extensive shocks in our simulations. 
Here, we attempt to disentangle the effects of the dynamics and the hydrodynamics
of the gas as it unbinds. We backtrack the unbound material at the 
end of the simulation through the snapshots to identify when the 
gas particle first unbinds. At the same time, we note when and if the gas particle  
is shocked. Table~\ref{table:shocked} shows the results of this process. We find
that for equal-mass mergers, galaxies on hyperbolic orbits systematically show that
a smaller fraction of the unbound material originates from shocks, compared to
galaxies on elliptical orbits. This may be because galaxies on
hyperbolic orbits only undergo one pericenter pass, while galaxies on elliptical
orbits experience multiple pericenter passes (and shocks). 
Elliptical orbits show a strong correlation between unbound and shocked particles. 
For instance, for equal-mass mergers, $\sim 85\%$ of 
all the unbound particles were shocked during the simulation. By tracking the particles back,
we find that the overwhelming majority of gas was unbound right after being shocked.

\begin{figure}
\centering
\includegraphics[scale=0.6,clip=true,bb=9 27 705 529]{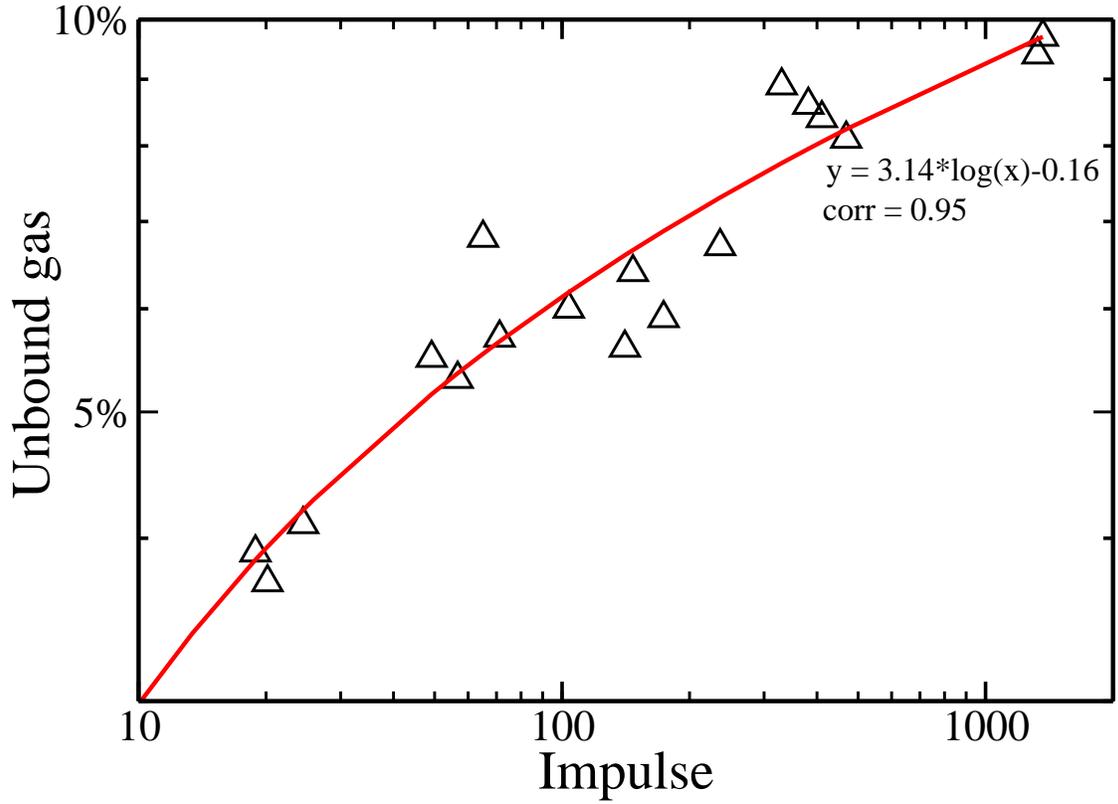}
\caption[The semi-analytic fit to the unbound fraction]
{\small The linear fit (red) for the unbound fraction and the impulse, 
$\Delta E$ (see Eqn.~\ref{eqn:shockimpulse}).
We only use the 18 
simulations with the 1:1 and 3:1 bound orbits since they leave a merger remnant in all cases
and have unbound fractions $> 3\%$. We will explore this in an accompanying paper to
find the unbound gas fraction throughout the history of structure formation.}
\label{figure:allunboundsemifit}
\end{figure}

In a similar spirit as the previous estimation of the peak $\LX$ from shocks, we 
quantify the effect of the orbit on the unbound material. We follow the formulation of \citet{CD08}
and use the impulse formulated in Eqn.~\ref{eqn:shockimpulse}. A linear regression 
analysis on $f_{\rm unb}$ and $\Delta E$ and obtain a linear relationship of the form: 
\begin{equation}
f_{\rm unb}  = 3.14\times\log_{10}\Delta E -0.16.
\label{eqn:allunboundsemifitform}
\end{equation}
The correlation coefficient for this fit is 0.95 (see Fig.~\ref{figure:allunboundsemifit}), implying that the two quantities are well
described by a linear function. Because we started out on a physically motivated basis by
equating the impulse with the unbound fraction, this high correlation points to 
an underlying relation between the impulse and the unbound fraction. 
Since most of the hyperbolic mergers do not leave a remnant and the 10:1 mergers have only $1\%$
unbound gas,, we do not use those simulations to fit this equation. Thus, out of our total
36 simulations, we have used only the 18  1:1 and 3:1 with elliptical orbits to obtain 
Eqn.~\ref{eqn:allunboundsemifitform}.

\clearpage
\newpage
\begin{landscape}
\renewcommand{\thefootnote}{\fnsymbol{footnote}}
\renewcommand{\arraystretch}{0.81}
\scriptsize

\begin{longtable}{cccccccccc}
\caption[Properties of the merger remnant for all the simulations]{Characteristics
of the merger simulation and the final remnant for all the simulations. Lack of 
an entry implies that the merger was not completed within a Hubble time.}
\label{table:mergers} \\

\hline \hline \\[-2ex]
   \multicolumn{1}{c}{\textbf{Gas Content}} &
   \multicolumn{1}{c}{\textbf{Peak L$\mathbf{_{\rm X}}$}\footnotemark[2]} &
   \multicolumn{1}{c}{\textbf{Peak Shock L$\mathbf{_{\rm X}}$}\footnotemark[3]} &
   \multicolumn{1}{c}{\textbf{Unb. gas mass}}  &
   \multicolumn{1}{c}{\textbf{Total unb. mass}} &
   \multicolumn{1}{c}{\textbf{Hot Gas}} &
   \multicolumn{1}{c}{\textbf{R$_{\rm \mathbf{vir}}$}} &
   \multicolumn{1}{c}{\textbf{Gas within \textbf{R$_{\rm \mathbf{vir}}$}}} &
   \multicolumn{1}{c}{\textbf{DM  within \textbf{R$_{\rm \mathbf{vir}}$}}} &
   \multicolumn{1}{c}{\textbf{Remnant L$_\mathbf{\rm X}$}} \\[0.8ex] 
   \multicolumn{1}{c}{\textbf{--}} &
   \multicolumn{1}{c}{\textbf{[$\mathbf{10^{40}}$ erg/s]}} &
   \multicolumn{1}{c}{\textbf{[$\mathbf{10^{38}}$ erg/s]}} &
   \multicolumn{1}{c}{$\mathbf{ [10^{10} \,\Msun ]}$} &
   \multicolumn{1}{c}{$\mathbf{ [10^{10} \,\Msun ]}$} &
   \multicolumn{1}{c}{\textbf{[\%]}} &
   \multicolumn{1}{c}{\textbf{[kpc]}} &
   \multicolumn{1}{c}{\textbf{[\%]}} &
   \multicolumn{1}{c}{\textbf{[\%]}} &
   \multicolumn{1}{c}{\textbf{[$\mathbf{10^{40}}$ erg/s]}} \\[0.5ex]\hline \hline \\[-2ex]
\endfirsthead

\multicolumn{8}{c}{{\tablename} \thetable{} -- Continued}\\[0.5ex]
  \hline \hline \\[-2ex]
   \multicolumn{1}{c}{\textbf{Gas Content}} &
   \multicolumn{1}{c}{\textbf{Peak L$\mathbf{_{\rm X}}$}} &
   \multicolumn{1}{c}{\textbf{Peak Shock L$\mathbf{_{\rm X}}$}} &
   \multicolumn{1}{c}{\textbf{Unb. gas mass}}  &
   \multicolumn{1}{c}{\textbf{Total unb. mass}} &
   \multicolumn{1}{c}{\textbf{Hot Gas}} &
   \multicolumn{1}{c}{\textbf{R$_{\rm \mathbf{vir}}$}} &
   \multicolumn{1}{c}{\textbf{Gas within \textbf{R$_{\rm \mathbf{vir}}$}}} &
   \multicolumn{1}{c}{\textbf{DM  within \textbf{R$_{\rm \mathbf{vir}}$}}} &
   \multicolumn{1}{c}{\textbf{Remnant L$_\mathbf{\rm X}$}} \\[0.8ex] 
   \multicolumn{1}{c}{\textbf{--}} &
   \multicolumn{1}{c}{\textbf{[$\mathbf{10^{40}}$ erg/s]}} &
   \multicolumn{1}{c}{\textbf{[$\mathbf{10^{38}}$ erg/s]}} &
   \multicolumn{1}{c}{$\mathbf{ [10^{10} \,\Msun ]}$} &
   \multicolumn{1}{c}{$\mathbf{ [10^{10} \,\Msun ]}$} &
   \multicolumn{1}{c}{\textbf{[\%]}} &
   \multicolumn{1}{c}{\textbf{[kpc]}} &
   \multicolumn{1}{c}{\textbf{[\%]}} &
   \multicolumn{1}{c}{\textbf{[\%]}} &
   \multicolumn{1}{c}{\textbf{[$\mathbf{10^{40}}$ erg/s]}} \\[0.5ex]\hline \hline \\[-2ex]
\endhead

\multicolumn{8}{c}{{Continued on Next Page\ldots}} \\
\endfoot

\hline\hline\\[-4.0ex] 
\endlastfoot
\multirow{12}{*}{\begin{sideways}{\Large{1\% gas}}\end{sideways}}
& 0.82   &  0.25   & 0.57 & 44.9 & 51.0 & --     &  --  & -- & 0.03\\
& 0.54   &  0.09   & 0.56 & 17.5 & 55.2 & --     &  --  & -- & 0.04\\
& 0.27   &  0.005  & 0.35 & 9.3  & 71.8 & --     &  --  & -- & 0.1\\
& 1.17   &  0.19   & 0.29 & 2.03 & 80.2 & 248.1  & 58.1 & 58.4 & 0.13\\
& 0.81   &  0.15   & 0.27 & 1.95 & 82.1 & 247.6  & 59.1 & 57.8 & 0.11\\
& 0.42   &  0.03   & 0.27 & 2.6  & 84.8 & 246.3  & 55.2 & 56.7 & 0.09\\
& 0.50   &  0.08   & 0.14 & 3.6  & 84.2 & 224.4  & 62.9 & 65.1 & 0.07\\
& 0.40   &  0.04   & 0.12 & 2.5  & 84.6 & 224.4  & 62.0 & 65.2 & 0.07\\
& 0.20   &  0.009  & 0.09 & 2.9  & 83.6 & 222.1  & 56.8 & 63.0 & 0.07\\
& 0.23   &  0.004  & 0.03 & 2.0  & 90.7 & 218.9  & 68.4 & 72.8 & 0.07\\
& 0.19   &  0.008  & 0.02 & 1.4  & 90.3 & 218.4  & 67.4 & 72.1 & 0.07\\
& 0.13   &  0.001  & 0.03 & 1.6  & 85.6 & 218.4  & 64.4 & 72.3 & 0.08\\[0.3ex]\hline\\[0.3ex]
\multirow{12}{*}{\begin{sideways}{\Large{10\% gas}}\end{sideways}}
& 90.5  & 14.4    & 7.4  & 33.9 & 52.6 & 217.5  & 25.5 & 39.1 & 3.0\\
& 41.7  &  6.0    & 6.1  & 24.5 & 52.9 & --     &  --  &  --  & 3.3\\
& 23.1  &  0.5    & 5.1  & 12.9 & 59.3 & --     &  --  &  --  & 9.0\\
& 79.3  &  18.7   & 3.05 & 4.36 & 79.4 & 251.3  & 53.8 & 60.4 & 12.5\\
& 71.9  &  10.2   & 2.84 & 4.16 & 80.9 & 248.6  & 53.3 & 58.3 & 9.5\\
& 34.5  &  1.7    & 2.06 & 3.86 & 83.3 & 246.7  & 49.8 & 57.1 & 8.1\\
& 44.0  &  7.6    & 1.4  &  4.2 & 82.2 & 224.4  & 56.0 & 64.9 & 6.2\\
& 36.0  &  3.4    & 1.2  &  3.3 & 83.6 & 225.3  & 55.3 & 65.8 & 6.8\\
& 18.9  &  0.4    & 0.84 &  2.8 & 84.3 & 223.9  & 52.2 & 64.6 & 5.6\\
& 26.5  &  0.4    & 0.22 &  2.0 & 85.2 & 216.6  & 58.7 & 70.4 & 5.5\\
& 16.4  &  0.2    & 0.22 &  1.4 & 88.4 & 217.1  & 57.7 & 71.0 & 5.8\\
& 11.2  &  0.1    & 0.23 &  1.6 & 84.5 & 215.7  & 55.1 & 69.7 & 6.4\\[0.3ex]\hline\\[0.3ex]
\multirow{12}{*}{\begin{sideways}{\Large{18\% gas}}\end{sideways}}
& 228.2 & 59.7    & 11.9  & 39.4  & 47.8 & 223.0  & 29.1 & 42.3 & 10.8 \\*
& 153.2 & 26.2    & 9.7   & 36.2  & 55.5 & --     & --   & --   & 10.3\\*
& 78.7  &  1.8    & 6.4   & 14.7  & 70.8 & --     & --   & --   & 27.6\\*
& 268.3 & 30.7    & 5.0   & 6.1   & 81.9 & 250.4  & 58.2 & 59.7 & 44.0\\*
& 227.4 & 26.3    & 4.7   & 5.8   & 83.1 & 251.8  & 57.8 & 60.6 & 35.0\\*
& 113.1 & 7.1     & 3.4   & 4.9   & 85.9 & 248.6  & 54.0 & 58.3 & 28.2\\*
& 193.5 & 20.8    & 2.3   &  4.7  & 85.6 & 224.8  & 59.5 & 65.3 & 22.2\\*
& 108.4 & 14.2    & 2.1   &  3.9  & 86.2 & 224.4  & 58.5 & 65.2 & 23.2\\*
& 61.8  & 1.8     & 1.6   &  3.9  & 86.7 & 223.9  & 54.6 & 64.6 & 19.6\\*
& 84.5  & 3.5     & 0.39  &  1.9  & 91.6 & 218.4  & 62.6 & 72.2 & 18.4\\*
& 54.8  & 1.7     & 0.38  &  1.6  & 90.8 & 216.6  & 60.7 & 70.4 & 19.3\\*
& 35.8  & 0.3     & 0.45  &  2.8  & 89.1 & 217.5  & 58.7 & 71.5 & 21.4\\*

\footnotetext[2]{L$_{\rm X}$ is calculated by using only the particles that have
$T_{\rm gas} > 10^{5.2}$ K  and $\rho < 0.01 $ M$_\odot /pc^{-3}$.}
\footnotetext[3]{Shock L$_{\rm X}$ is obtained by adding the L$_{\rm X}$ for
the particles that exceed a threshold for $\frac{\dd S}{\dd t}$. The threshold,
in code units, is $10^{10}, 3\times10^9$ and $2\times10^9$ for $1\%, 10\%$
and $18\%$ gas fractions.}
\end{longtable}

\normalsize
\renewcommand{\thefootnote}{\arabic{footnote}}
\renewcommand{\arraystretch}{1.0}
\end{landscape}

\section{DISCUSSION}\label{section:discussion}

In our simulations, we found that the hot halo gas shock-heats to temperatures
$\sim 10^{6.3}$ K in equal-mass mergers and $\sim 10^{6}$ K in unequal-mass
mergers. This is reflected in a strong temperature jump in the regions between
the two colliding galaxies, well before pericenter passage, and persists until the
first pass. The strongest shocks, and correspondingly the largest X-ray luminosities
due to shocked gas, are created after the pericenter passage. For 
gas fractions greater than 10\%, 
this $\LX$ from shocks remains above observable thresholds of  $\sim 10^{39}$ erg/s
for at least a period of $\sim 300$ Myrs, and thus should be detectable in  
ongoing mergers with {\it Chandra} and {\it XMM-Newton}. Compared to the merger 
timescale, an observable X-ray shock is $\sim 20$ times shorter.
During the merger, the hot gas itself also radiates and can have an X-ray luminosity
of $\sim 10^{42}$ erg/s  for galaxies with a gas fraction as low as 10\%.
Herein lies the difficulty of detecting the shocked gas: the shocked gas is 1000 times
less luminous than the non-shocked gas. To observe such X-ray shocks from mergers, one 
would have to look for a signal 3 orders of magnitude smaller
than the global X-ray level {\em {and}} the effect only lasts for a timescale
that is an order of magnitude shorter than the merger. Merger-induced star formation and AGN activity adds to the confusion 
by contributing X-rays from accretion disks. 

The redeeming feature of our predictions comes from the fact that the
X-ray emission occurs even when the galaxy separation is large and occurs in a region where the X-ray
background is low. Therefore, the ideal procedure 
to validate our predictions would be to observe close pairs of galaxies with an undisturbed
morphology (indicating that the first pass has not occurred), with current
space telescopes like {\it Chandra} and {\it XMM-Newton}. In such a case,
there will be no contamination from merger-associated starbursts and the $\LX$
signature, if observed, could be uniquely ascribed to shocked halo gas.
Our simulations assume that the gas is always heated to the virial temperature
in galaxies, irrespective of the galaxy mass; recent numerical simulations
predict that that may not be the case \citep{KKWD05}. Smaller mass galaxies, similar to the 
smallest galaxy we simulated (with $v_{\rm circ} = 74$ km/s), are not expected to 
heat above $10^{4.5}$ K. In such a scenario, the predicted $\LX$ from shocks
will be much lower. We can use this prediction to constrain the theories of galaxy formation 
(both semi-analytic and from numerical simulations) against actual observations. 

A detailed analysis of the feasibility of the X-ray detection is beyond the scope of this
paper; however, we will assume a conservative threshold of the X-ray flux as 
$10^{-15}$ erg/s/cm$^2$ and some representative distance to galaxy pairs 
of $\sim$ 20 Mpc. This sets a detection threshold in the X-ray 
luminosity of $5\times10^{37}$ erg/s. We will find the hot halo itself exceeds this threshold 
for all the simulations  but the shock X-ray luminosity does not always do so.
We empirically model the peak $\LX$ from shocks based on our set of 36 simulations as follows:
\begin{equation}
\log_{10}\,\left[\frac{L_{\rm peak}}{10^{38} \;{\rm erg/s}}\right] = 0.88 \times \log_{10}\,\Delta H - 1.25.
\end{equation}
Note that we have modelled $\LX$ completely by thermal \brems, which is a fairly
inefficient radiation mechanism at $\sim 10^{6.2}$ K. Using {\it XSPEC}, we find if that gas is at solar
metallicity, the total emission will increase by more
than an order of magnitude. 
Thus, our predictions about $\LX$ should be considered a lower limit
to the likely X-ray emission in actual observations of colliding galaxies for a given hot gas mass. 

Our simulations show that gas escapes from  galaxies during the course of the merger,
with most of the gas propelled outside 3 $\Rvir$ even before the final merger event. 
For equal-mass mergers, the unbound gas fraction ranges from 10-20\% depending on the 
impact parameter and orbit type. The unequal-mass
mergers result in 3-5\% unbound gas, with similar fractions coming from each of
the merging galaxies. All of this unbound gas, presumably enriched
with metals, flows out into the IGM. 
Observations of the IGM at low redshift show the presence of metals \citep{DS08}; cosmological simulations
can account for the metals since it only takes trace amounts of metals to 
leave a large-scale signature. However, about 50\% of the baryons are actually present 
in the IGM \citep{FP04} in a temperature-density phase-space that is difficult to probe via observations \citep{DS08}. 
For about a decade, numerical simulations of
structure formation have been struggling to obtain a prescription that allows 
such large amounts of gas to escape from galaxies without violating other observational
constraints like the star formation rate. Our results show that mergers are an effective way to extract gravitational 
energy and cause significant  mass-loss from galaxies. 
This process is similar, in principle, to the gravitational heating that has been incorporated for the first time in semi-analytic
galaxy formation recipes in a recent paper \citep{KO08}. Our simulations provide
the empirical fits that can be incorporated into such semi-analytic simulations.
Though the amount of gas has a strong dependence on the initial orbital energies, we  see that each merger
releases $\gtrsim 2\times10^{10}\,\Msun$ of material into the IGM. 
Given that each galaxy suffers a major merger in its past \citep{CBDP03}, the total amount 
of unbound gas can become significant over the lifetime of the Universe. We calculate this
in a companion paper \citep{SH09}.

We use those merger remnants with unbound fractions larger than 3\% to obtain a prescription for the unbound gas fraction
as a function of the merger impulse, $\Delta E$, as follows:
\begin{equation}
f_{\rm unb}  = 3.14\times\log_{10}\Delta E -0.16.
\end{equation}
Overall,  both X-ray shocks and the unbound gas production 
are two facets of the same phenomena: the conversion of gravitational potential energy
into thermal and kinetic energy. Even though cooling, star formation and other
baryonic physics were not included in our simulations, the tremendous amount of 
available gravitational potential energy and the subsequent conversion of a fraction
of it into  thermal energy continuously heats the gas at 
a much faster rate than it can cool via any feasible cooling mechanism, at least for the 
temperatures and densities involved in our simulations.

\citet{CPJS04} showed that it is possible 
to repopulate the halo with gas (and metals) via mergers of disk galaxies; mass-loaded 
stellar winds from high mass stars, and AGN activity can also enrich the gas in the halo. 
We know that the metal content of the gas can strongly enhance the X-ray production,
even overshadowing the contribution from thermal \brems. For example, 
for solar metallicity gas at $10^6$ K, the emission for a {\small MEKAL} model is 2 orders of magnitude larger than that of 
thermal \brems.  As part of a future project, we will
include metals in the merging galaxies to determine how this affects the  X-ray
properties. We expect that the boost from metal line emission will make
the emission more detectable, and may allow X-ray shocks to be observed for 
galaxy pairs with larger separations.

Our simulations were performed with vacuum boundary conditions, but in reality
galaxies are continuously evolving objects within the background of an expanding 
Universe. Pristine gas is falling into the haloes, and the haloes themselves are
growing more massive. Cosmological simulations are required to fully model the
mergers within the context of an evolving background. As part of a future
project, we will use cosmological simulations with dark matter and gas to
explore the phenomena of the shocks in the mergers. 

An important implication of this work is that using our fitting formulae we can estimate the total
emission in X-rays from shocks and the total amount of 
unbound gas as a function of $z$ over the history of the Universe. 
To do this, we will combine  an extended Press-Schechter formalism 
\citep{LC93,B91,BCEK91} with our empirical fits, in a semi-analytic approach. 
The semi-analytic estimate for the unbound gas fraction over the lifetime of the Universe
can be compared against the current gas fractions of galaxies as well as the baryonic content of
the IGM. We will explore the evolution of the unbound gas fraction of the Universe in \citet{SH09}.

\clearpage
\bibliographystyle{apj}
\bibliography{Biblio-Database}

\end{document}